\documentclass[OPT,biber]{nowfnt}
\usepackage[utf8]{inputenc}
\usepackage{graphicx}
\usepackage{hyperref}
\usepackage{fancyhdr}
\usepackage{xcolor}
\usepackage{acronym}
\usepackage{enumerate}
\usepackage[inline,shortlabels]{enumitem}
\usepackage{bigstrut}
\usepackage{rotating}
\usepackage{amsmath}
\setlist[enumerate]{noitemsep}

\newcommand{\revision}[1]{\textcolor{black}{#1}}

\usepackage{soul}

\title{Domain-specific Information Retrieval}

\maintitleauthorlist{
Suzan Verberne \\
Leiden University \\
s.verberne@liacs.leidenuniv.nl
\and
Florina Piroi \\
TU Wien \\
florina.piroi@tuwien.ac.at
\and
Allan Hanbury \\
TU Wien, Complexity Science Hub\\
allan.hanbury@tuwien.ac.at
}

\issuesetup
{%
 copyrightowner={S.~Verberne and F.~Piroi and A. Hanbury},
 volume        = xx,
 issue         = xx,
 pubyear       = 2026,
 isbn          = xxx-x-xxxxx-xxx-x,
 eisbn         = xxx-x-xxxxx-xxx-x,
 doi           = 10.1561/XXXXXXXXX,
 firstpage     = 1, 
 lastpage      = 130
 }

\usepackage{mwe}

\author[1]{Verberne, Suzan}
\author[2]{Piroi, Florina}
\author[3]{Hanbury, Allan}

\affil[1]{Leiden University; s.verberne@liacs.leidenuniv.nl}
\affil[2]{TU Wien; florina.piroi@tuwien.ac.at}
\affil[3]{TU Wien, Complexity Science Hub; allan.hanbury@tuwien.ac.at}

\begin{document}

\maketitle
\section*{Abstract}
Information Retrieval (IR) is much more than ad hoc web search: a large portion of the information seeking activities takes place in specific domains, often addressing work-related tasks. Examples are lawyers finding relevant prior cases for a legal case and academic researchers studying the background material for a paper. In Chapter 2 we outline the most common domain-specific tasks, datasets and applications. Domain-specific IR has a number of challenges that make set it apart from simple web search tasks: they often deal with long documents, long queries and long sessions with specific user needs. We discuss these in Chapter 3. We also address issues related to bias and fairness of models. In Chapter 4, we describe the models and methods developed for domain-specific tasks and data, addressing the specific challenges, including user aspects related to domain-specific IR tasks, dealing with queries, user models, and user interactions. We show that many of the methodological advances in domain-specific IR are transferable to other domains and even to open-domain web search. We conclude with the main takeaways and propose research directions for the near future.


\chapter{Introduction} \label{chap:intro}
The ubiquity of Web Search has ensured that the dominant search paradigm is a query box through which a query is submitted by a user to a search engine accessing an index over a huge number of documents on multiple topics. The results of this search are displayed as a ranked list of search results. While this paradigm is suitable for a large variety of search tasks, for domain-specific, specialised search tasks more comprehensive tools are required. 

This is particularly true in the area of {\em professional search} \citep{russell-rose_information_2018,verberne2019first}. Professional search focuses on the work of paid professionals who are undertaking a work task that is predominately search-related and performed under a number of constraints such as budget and time \citep{Tai2014}. Professional search covers a wide range of domains, tasks, and tools. At one end of the range, a professional search task could last several hours to days, involve a formal search language (e.g. Boolean search), and have potential legal repercussions if relevant documents are not found. This applies to prior art patent search \citep{LH13} as well as prior case search in case law legal systems \citep{locke2022case}. Search for publications on clinical trial outcomes as the basis for a systematic review in the medical domain also last several hours to days, use Boolean search, but missed documents could lead to an incorrect conclusion about a medical intervention \citep{KZK23}, resulting in potentially less effective treatments being used on patients.

At the other end of the range are search tasks that can best be solved by simply applying the web search paradigm using an index that is restricted to a certain set of documents, \revision{also referred to as {\em Vertical Search} \citep{Lewandowski2023}}. This \revision{index restriction} has also been used as a simple definition of \textit{domain-specific search}, e.g., by \citet{TCH06}. Google has created usable generic tools for patent\footnote{\url{https://patents.google.com/}} and academic paper search\footnote{\url{https://scholar.google.com/}} by essentially using its web search approach restricted to patents and academic papers. While these tools are far removed from satisfying the requirements for the professional search tasks mentioned in the previous paragraph, they are sufficient, for example, for a scientist to get a rapid overview of the state-of-the-art in a narrow area, or find patents or papers useful for solving a problem being faced. \revision{In professional contexts, the documents to be searched can be the same as in generic web search, but may be extended with limited-access content such as intranets or restricted databases.}

Domain-specific search can also be conducted in a non-professional context. For example, patients often search for medical information on the internet, or a university student may search for options provided by the law to extend the maximum study time due to illness, so as to avoid losing tax benefits. Such searchers often need support in other areas, such as assistance in determining the credibility of the information found.

\section{Model of Domain-Specific Search}\label{sec:model}

Domain-specific search can be more complex than simply restricting the documents to be indexed to a set of documents relevant to the domain, as illustrated above. In this section we present a model of domain-specific search, building on existing models, that attempts to make explicit all aspects that should be considered when designing and implementing a domain-specific search solution.

\citet{HL13} introduced the following five-dimensional model for domain-specific search:
\begin{quote}
    A \textbf{domain-specific search [engine $|$ process]} is a search [engine $|$ process] that specifies each of the following five dimensions: 
\begin{enumerate}
\item \textbf{subject areas} e.g. chemical, biomedical, healthcare 
\item \textbf{modality} e.g. text, images, videos, audio
\item \textbf{users} e.g. a patent examiner, a professor of medicine, a project manager 
\item \textbf{tasks} e.g. prior art patent search, technology survey, literature search, systematic review, diagnosis search
\item \textbf{tools, techniques and algorithms required} to complete the tasks, e.g. query completion limited to specific vocabularies, cross-lingual search, possibility to store search results
\end{enumerate}
\end{quote}

\begin{figure}[tb]
    \centering
    \includegraphics[width=0.5\textwidth,alt=The design triangle for domain-specific search]{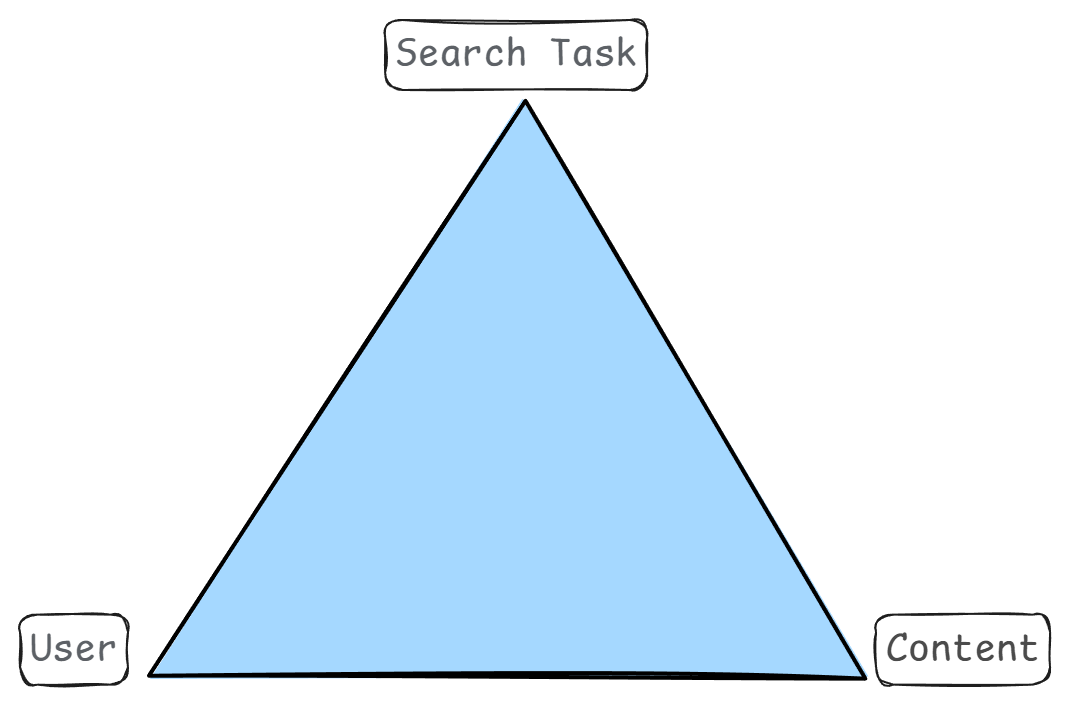}
    \caption{The design triangle for domain-specific search, inspired by \citet{MA14}.}
    \label{fig:design_triangle}
\end{figure}

\revision{The dependencies between the dimensions has been, later, discussed by \citet{LSH14}}. For example, modality is highly correlated with subject area, while the tools, techniques and algorithms depend on the specifications of the first four dimensions. At around the same time, \citet{MA14} introduced the {\textbf{design triangle}} in the area of visual analytics. The design triangle emphasizes the role of {\em data}, {\em users} and {\em tasks} in designing visual analytics solutions. Once these aspects have been specified, decisions can be made on selecting the most appropriate tools, techniques and algorithms to use. The design triangle can straightforwardly be adopted to domain-specific search (by replacing \textit{data} with \textit{content}), as shown in Figure~\ref{fig:design_triangle}. 
This is a reduction of the five-dimensional model, where the {\em content} aspect groups the {\em subject areas} and {\em modality} dimensions, while {\em users} and {\em tasks} map directly to each other. In this model, the {\em tools, techniques and algorithms required} are not seen as a dimension of the model, but as a specification of the solution to be built based on the specification of the model. This model allows a rapid characterisation of domain-specific search as triplets of $$ [user,\ task,\ content]$$ such as $$ [\mathrm{patent\ examiner,\ patentability\ search,\ scientific\ and\ patent\ literature}]$$ 
or 
\begin{multline*}
[ \mathrm{medical\ doctor,\ finding\ a\ clinical\ trial\ in\ which\ to\ enroll\ a\ patient,} \\
\mathrm{database\ of\ clinical\ trials\ currently\ recruiting} ]
\end{multline*}

One of the uses of such a model is to facilitate the transfer of search solutions between domains. If the difference between domains is characterised only by the  content indexed, then this could lead to a new solution being developed in relative isolation whenever sufficiently different content is encountered \citep{HL13}. The consideration of users and tasks in the model allows further similarities between domains to be identified and insights about similar tasks and users to be transferred between domains with different content. \citet{russell-rose_information_2018} illustrate this through an online survey carried out among professional searchers in multiple domains: legal researchers, recruitment professionals, and healthcare information professionals. This supplements the results of a similar survey carried out among patent professionals by \citet{JAV10}. Similarities between the users in all domains were found to be a preference for using Boolean queries; the need to manage, organise and re-use search strategies and results; and an ambivalence toward the use of relevance ranking. An identified difference was a precision-orientation in the legal and recruiting domains compared to a recall-orientation in the patent and medical domains. Such insights into users and tasks already indicate IR techniques and approaches that can be transferred between domains. For example, \citet{SR22} investigate visual Boolean query builders applicable to multiple domains.

The rapid characterisation of a domain in terms of the brief descriptions of each of the three aspects does provide a good starting point in specifying a solution and contextualising solutions in the solution space. However, each of the aspects represents a wealth of detail that is abstracted away in this representation. We now dig deeper into the detail of what each of the aspects represents and what the implications of each of the aspects on domain-specific search solutions are. 
\revision{Both \citet{Tom11} and \citet{ingwersen_turn:_2005} examine} in detail the definition and characteristics of search {\em tasks}, as well as how to model them and place them in the context of broader {\em work tasks}. \revision{\citet{bystrom1995task} and \citet{bystrom2005conceptual} also gave the task a central role in their model. In their 1995 work, their information seeking model (building on earlier work from the 1970s and 1980s) puts the information need analysis first, with input from personal factors, the subjective task, situational factors, and personal style. After the information need analysis the model follows with the choice of action, implementation, and evaluation. In their 2005 paper, \citet{bystrom2005conceptual} define three task phases: task construction, task performance, and task completion. The phases are inter-connected and not linearly executed by the user.}

\begin{figure}[t]
    \centering
    \includegraphics[width=1.0\textwidth,alt=The context of a search task]{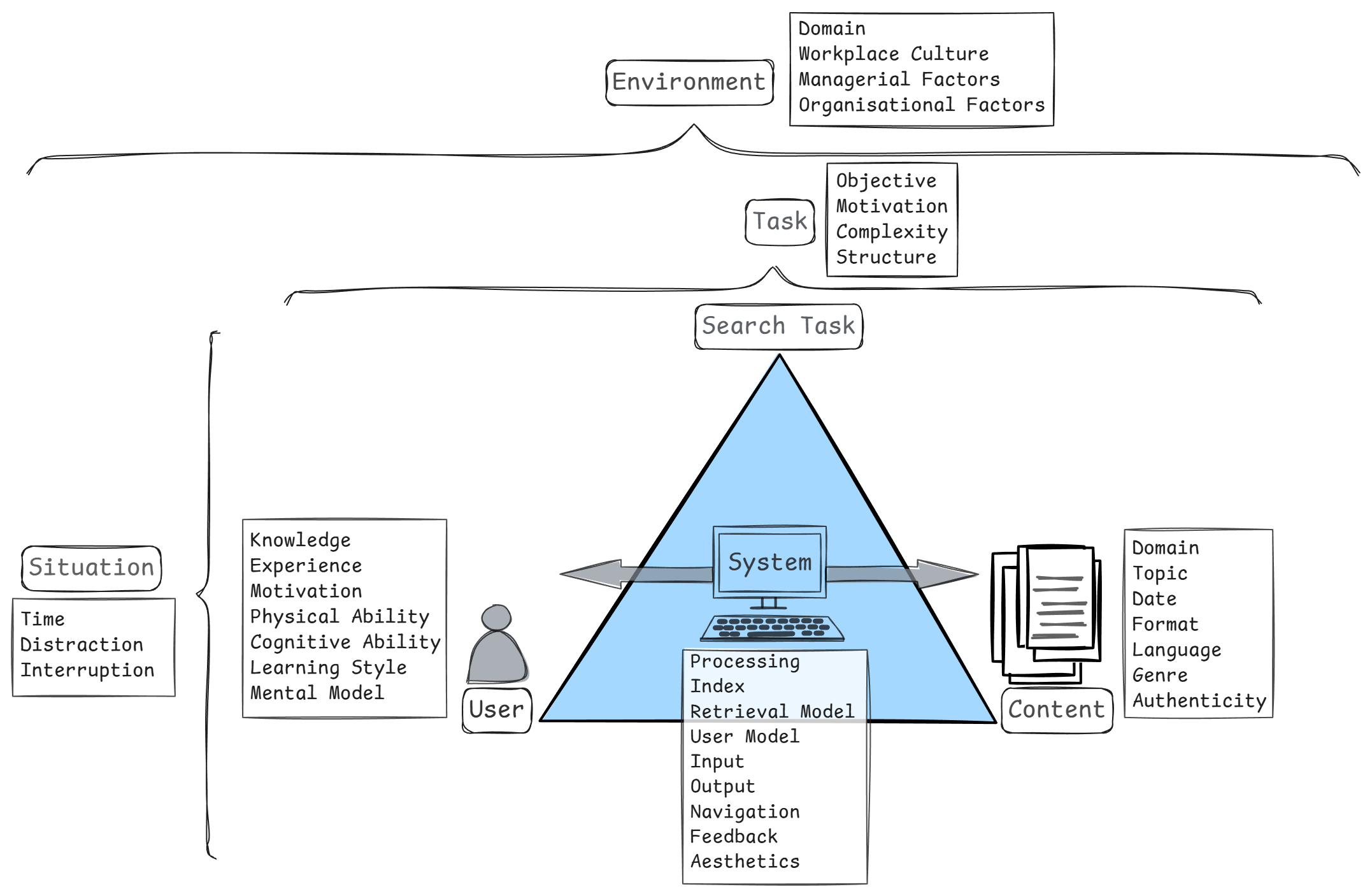}
    \caption{The context of a search task, inspired by \citet{Tom11}. The aspects and broader contexts are shown as orange rectangles, while characteristics of these aspects and contexts are listed in their adjacent white rectangles. The blue triangle in the centre corresponds to the triangle in Figure~\ref{fig:design_triangle}.}
    \label{fig:task_context}
\end{figure}

\begin{figure}[t]
\includegraphics[width=0.9\textwidth]{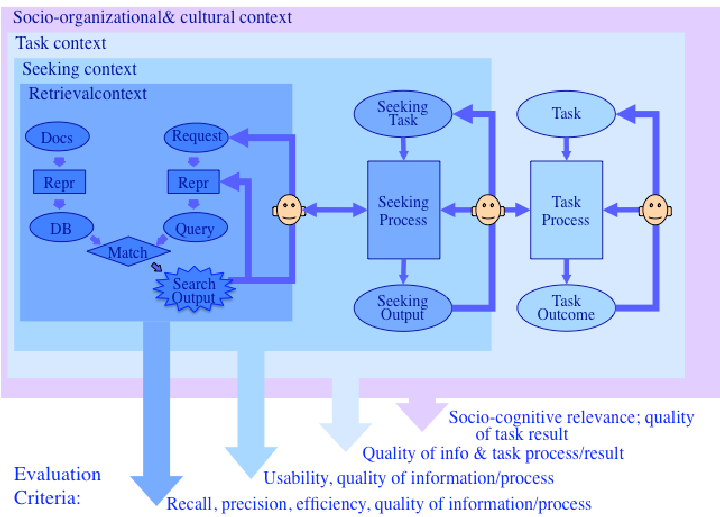}
\caption{The model of nested evaluation contexts in task-based information seeking \citep{ingwersen_turn:_2005}}\label{fig:ingwersencontext}

\end{figure}

Figure~\ref{fig:task_context} shows a diagram, adapted from \citet{Tom11}, which breaks \revision{each of the three aspects in the design triangle down to further aspects and components to be considered}. The blue triangle of Figure~\ref{fig:design_triangle} is visible in the centre of the diagram, linking the three aspects with the search system in the centre of the triangle. Here again, the search system is the outcome of the specification of the three aspects corresponding to the triangle vertices --- these aspects help to select the characteristics of the search system, listed below the ``System'' rectangle in the figure. \revision{Figure \ref{fig:ingwersencontext} shows the model of \citet{ingwersen_turn:_2005}, placing retrieval in the context of information seeking, and information seeking in the context of a task.}

\revision{In both diagrams}, the separation between {\em task} and {\em search/seeking task} is made explicit. The search task is usually conducted as part of a broader task that involves additional steps, such as analysis and documentation. For example, the task could have the objective of the creation of a systematic review of a medical intervention. Some of the steps of the task are search tasks, searching for relevant medical literature, but other steps include writing a protocol, analysing the search results, extracting information from the relevant papers, performing a statistical analysis, and writing up the outcome. In order to be reproducible, the task of creating a systematic review in medicine is strictly standardised (see Section~\ref{sec:health}), which fixes the structure and complexity of the task. For example, the Boolean queries used to retrieve the documents must be published as part of the systematic review. Deviations from this standardised workflow are not permitted if the systematic review is to be accepted by the community. This strict standardisation is one of the organisational factors of the {\em Environment} in which the task is carried out. Another environmental factor could be the amount of progress reporting to management that the person performing the task is required to do, which could affect how efficiently a task is carried out. Search tasks are present both in professional search \revision{-- in which case the processes to carry them out are often clearly defined -- and as part of ``everyday'' tasks, carried out using a web search, such as such as organising a vacation, as described by \citet{Whi24}.}

The characteristics of the {\em User} carrying out the task can vary significantly and have an effect on the type of search and search interface that the user is capable of using or prepared to use and the documents they need to find. For example, the mental model of a software system held by a senior developer of the system \revision{is likely to be significantly different from the mental model of the same system that} a call-centre employee providing first-level technical support holds. Although both will search through the documentation of the system as part of their employment, they have widely different knowledge and experience leading to differences in how they would interpret the same documents. The situation in which the users work can also affect how they search. \revision{For example,} working with frequent distractions and interruptions may limit the complexity of the search task that is possible.

Lastly, the characteristics of the {\em Content} to be searched range from all of the content on the web to content restricted to specific academic, government\revision{, or internal company} sources, with other possible restrictions by e.g. date or language. Privacy and security aspects could lead to an interaction between the {\em User} and {\em Content} aspects, as for some tasks, users may have differing access rights to view documents. 

Obtaining a deeper understanding of these factors is usually best done through interaction between the Information Retrieval experts and the domain practitioners, through techniques including questionnaires~\citep{verberne2019information}, observation and discussion. As an illustration of the complexity of analysing a search task, we present in Figure~\ref{tab:DS_template} a task analysis template \citep{DoSSIER_D1_1} showing the diverse aspects that characterise a single domain-specific search task.

\begin{figure}
\caption{\label{tab:DS_template}Task analysis template for domain specific search}
\noindent\fbox{%
    \parbox{\textwidth}{%
    \small
\begin{description}
\item[task name:]{\em the unique name assigned to the task, e.g. Prefiling Patentability Search}

\item[definition:]{\em a brief definition of the task}
 
\item[rationale:]{\em why is the task carried out? what should carrying out the task achieve? e.g. the task should lead to the identification of one or more patents that invalidate the query patent.}

\item[initial information available:]{\em what information is available at the beginning to start the search? e.g. a patent application document}

\item[information source:]{\em what information must be searched? e.g. all patent and non-patent information published prior to today.}

\item[searcher:]{\em who usually performs the search? e.g. subject expert or librarian}

\item[query formulation methodology:]{\em how are the queries formulated? e.g. extraction of keywords from the query document and formulation of a Boolean query using synonym expansion lists}

\item[types of tools used:]{\em what tools are commonly used in this type of search? e.g. clustering results, merging results, Boolean search, ...}

\item[search stopping criteria:]{\em what criteria are used to decide when the search process must be stopped? e.g. a reasonable number of documents returned by a Boolean query}

\item[output of the search:]{\em what does the result list look like? e.g. a list of patents matching the Boolean query in reverse chronological order.}

\item[how/if the search is documented:]{\em is the search documented in some standard way? e.g. queries are placed into a search report along with the number of documents retrieved per query.}

\item[post-processing, interpretation, and analysis of search results:]{\em what is done with the result list once it is obtained? e.g. every patent is checked for relevance by an expert, if relevant it is marked as X or Y...}

\item[any caveats to consider in the analysis or its interpretation:]{\em e.g. the searcher needs to have a good understanding of what the requestor is looking for to enable a quick review of the answers for relevance.}

\end{description}
    }
}
\end{figure}

\section{Model Abstraction}
\label{subsec:model-abstraction}
\begin{figure}[t]
    \centering
    \includegraphics[width=1.0\textwidth,alt=The overview of the abstraction of the domain-specific search model.]{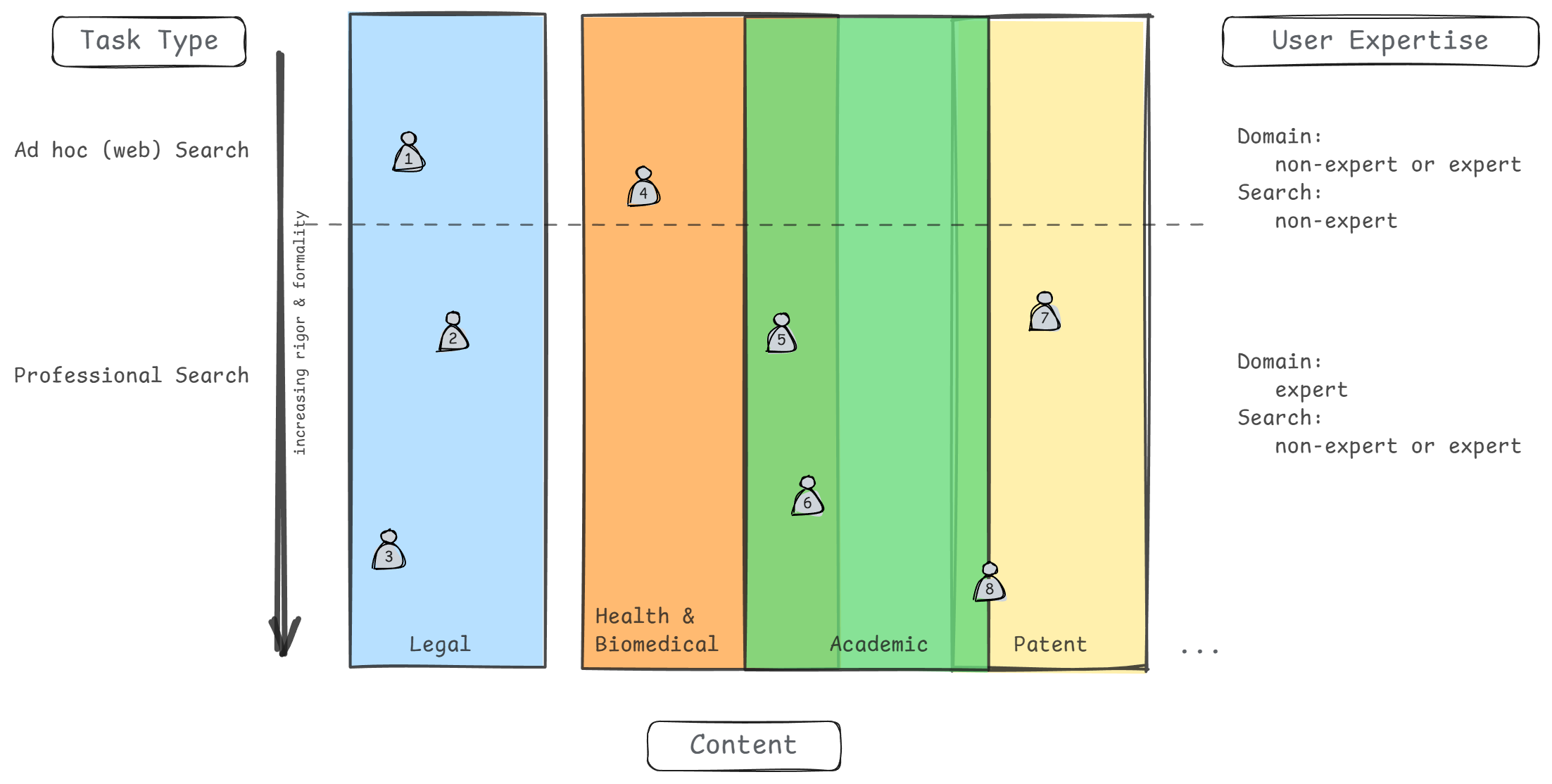}
    \caption{An abstraction of the content, task and user domain-specific search model.}
    \label{fig:overview_diagram}
\end{figure}

The design triangle model of domain-specific search is flexible and allows each of the aspects to be considered in depth, as illustrated above. This flexibility is useful for designing search solutions and systems, but it could hinder the use of the model to compare systems and solutions to each other. We therefore make an abstraction of the design triangle for domain-specific search (Figure~\ref{fig:design_triangle}), as shown in Figure~\ref{fig:overview_diagram}.

    In Figure~\ref{fig:overview_diagram}, the $x$-axis abstracts the domain aspect of the content, with each vertical bar representing the documents within a domain. Overlapping bars indicate that some documents can be considered in multiple domains. 
    The $y$-axis abstracts the tasks into two task types: \revision{Ad-hoc (web)} search and Professional search, \revision{with no clear border between them, which we mark with a dashed horizontal line}. The much wider range of tasks within professional search (mentioned by~\citet{russell-rose_information_2018}) compared to the range of ad-hoc search tasks is reflected in the figure by the level of the horizontal dashed line. 
    The user \revision{expertise} is abstracted \revision{along two dimensions: domain expertise and search expertise. Ad-hoc web search can be conducted by both domain experts and domain non-experts, while professional search is always conducted by domain experts. Search expertise relates to the professionalism of the search process: information professionals have the expertise to conduct an extensive (often complete) search for a certain topic.} 
The diagram underlines that we see professional search as a subset of domain specific search, but that domain specific search also includes search by non-experts\revision{, while some professional search entails search expertise in addition to domain expertise}.

The advantage of the diagram is that we can straightforwardly visualise common domain-specific search tasks in terms of their task complexity, user and content on 2D axes. To illustrate this, the numbered \revision{user icons} on the diagram correspond to the following concrete examples of users, tasks, and content:
\begin{enumerate}
\item A medical doctor conducting a web search to find the legal requirements for flying a drone equipped with a camera in a specific country.
\item A lawyer searching for a known case so as to include the correct case reference in a legal opinion she is writing.
\item A paralegal conducting a search of all internal documents of a company as part of an E-Discovery \citep{oard2013information} process.
\item An electrical engineer searching for information on mitigating the effects of a chronic disease of a member of his family.
\item A General Practitioner (GP) searching for treatment guidelines for a less common disease afflicting a patient.
\item An information professional carrying out a systematic review on the efficacy of a certain intervention for a specific disease.
\item A scientist at an industrial research laboratory conducting a search for existing patents that are obviously infringed by her invention before submitting a description of the invention to the Intellectual Property department.
\item A patent examiner at a national patent office searching for prior art related to a specific patent application.
\end{enumerate}

\section{Scope of the work}\label{sec:scope}

This paper focuses on research on domain-specific search tasks in the IR community in the last 15 years. These tasks are primarily identified by the organisation of evaluation campaigns within the main IR evaluation venues (TREC,\footnote{\url{https://trec.nist.gov/}} CLEF,\footnote{\url{https://www.clef-initiative.eu/}} NTCIR,\footnote{\url{https://research.nii.ac.jp/ntcir/index-en.html}} FIRE\footnote{\url{https://fire.irsi.org.in/}}) or the release of well-adopted datasets for the tasks. We focus on tasks that involve only text documents, but briefly reference work on tasks using multimodal data. While Enterprise Search \citep{Haw10} also satisfies the definition of domain-specific search, it is not included due to a lack of recent evaluation or shared tasks. 
We focus on tasks for which the retrieval unit is either a document or part of a document (paragraph, sentence). 

\revision{The current search engine landscape shows an evident move towards LLM-driven conversational interfaces (i.e. Question Answering--QA). As these engines necessarily include retrieval components at their core, we mention only the most relevant research on (the retrieval aspects of) QA, and do not cover all work on domain-specific QA.} 
\revision{Also note that the focus of our survey is on domains, challenges, and algorithmic (`back end') solutions, and less on user interface design and user behaviour aspects. We briefly discuss query interfaces in Section \ref{sec:queryproc}~(\nameref{sec:queryproc}) and conversational interfaces in Section~\ref{sec:genLLMs}~(\nameref{sec:genLLMs}), but refrain from an extensive discussion of literature from the human--computer interaction context. For this topic we refer to \citet{liu2021search,wilson2022search}.}

In this paper, we focus on the following domains: the legal domain, the health and biomedical domain, the academic domain, and the patent domain. All these have a profound socioeconomic impact on society and search in these domains has played a central role for at least 50 years. They have large communities of professional searchers who have developed and adopted codes of practice for search. However, for each of these domains there are many users conducting ``informal'' domain-specific search tasks that do not follow any codes of practice.
To understand the needs and requirements of the users of domain\nobreakdash-specific search, the European Union funded 15 PhD theses on domain\nobreakdash-specific search within these domains\footnote{With academic and patent search merged into innovation search} as part of the DoSSIER Marie Skłodowska\nobreakdash-Curie Innovative Training Network\footnote{DoSSIER: Domain Specific Systems for Information Extraction and Retrieval, \url{https://doi.org/10.3030/860721}} (2019--2024). Consequently there has been an increase in research and hence publications on these domains during this time. This monograph is, to a large extent, inspired by the research carried out in the DoSSIER project. 

We also specify that we concentrate on information retrieval tasks --- even though there are multiple Natural Language Processing (NLP) and Information Extraction (IE) tasks that are related to domain-specific search --- we do not include them in this overview. 

Finally, we focus mainly on English as the language of domain\nobreakdash-specific search, following the predominance of English in IR research. We acknowledge that research on other languages than English is important, and we discuss some of this in the context of legal IR in Section~\ref{sec:legal} and patent retrieval in Section~\ref{sec:patents}. We do think that research addressing non-English domain-specific IR is an important topic for future work (also see Section~\ref{sec:limitations}~\nameref{sec:limitations}).

\section{Work in the IR community on domain specific search}

\citet{KG01} present the advantages of evaluating on a domain\nobreakdash-specific test collection, in their case a test collection of social science documents in German (GIRT -- German Information Retrieval Test database). The main challenge described is the vocabulary mismatch between such a domain\nobreakdash-specific test collection and a more generic test collection, such as newswire and newspaper articles. An additional challenge \revision{identified was} that the \revision{(retrieval)} topic developers and the assessors must be familiar with the domain. Further early domain\nobreakdash-specific tasks were \revision{organized at} TREC-7 and TREC-8, which used a previous version of the GIRT test collection.\footnote{This was of course not the first domain-specific test collection for IR -- the first test collection created for IR evaluation by Cleverdon was specific to the academic domain, consisting of aerospace-related abstracts \citep{Rob08}.}

The majority of research on domain-specific IR has been driven by relevant tracks at the main IR evaluation venues: CLEF, TREC, NTCIR and FIRE. Through these tracks, researchers get access to test collections for search tasks, often linked to specific \revision{challenges faced at the time of the collection releases, many of those challenges still being relevant.}

Domain\nobreakdash-specific evaluation tracks in these venues are shown in Table~\ref{tab:evaluation_tracks}, starting in the year 2000. We only list those tracks that have at least one task that is clearly search-focused and do not include the tracks that are purely NLP or IE. For TREC, CLEF and FIRE, a year is indicated if a workshop for the corresponding track took place in that year. Due to \revision{the NTCIR 18-month cycle, its workshops are} not held every year. Hence, for the NTCIR, a bar is drawn for a corresponding track between the first workshop and final workshop including that track. In order to be able to follow the development per domain, the tracks have been clustered by domain, and sorted within the domains by the starting year of the track. 

Each of the domains covered in this paper are characterised by an early and relatively long-running track: the CLEF {\em Domain-Specific Retrieval} in the Academic domain, the NTCIR {\em Patent Retrieval/Mining} track in the patent domain, the TREC {\em Genomics} track in the health and biomedical domain, and the TREC {\em Legal} track in the legal domain. \revision{An additional competition in the legal domain is CoLIEE: the Competition on Legal Information Extraction and Entailment. CoLIEE aimed to create a community of practice in legal informatics. Since 2018 it includes a task on Legal Information Retrieval.\footnote{\url{https://coliee.org/resources}}}

Since 2014 there have been no evaluation tracks in the patent domain, and only short-running tracks in the academic domain. The health and biomedical domain has been represented by many evaluation tracks covering various aspects from search by patients to systematic reviews, and continues to have multiple evaluation tracks in recent years. While search in the mathematics domain could also be seen as part of academic search, we show these tracks separately in the table due to the specific retrieval techniques needed for mathematics retrieval. Three tracks focussed on multiple domains: the TREC {\em Chemical IR} included search of patents and academic publications; the TREC {\em Total Recall} Track focussed on E-Discovery (E-Mail search) in the legal domain, but also included news, Twitter, medical, and E-Commerce datasets. Similarly, the TREC {\em Dynamic Domain} Track included medical, E-Commerce, and news datasets.

\begin{sidewaystable}[htbp]
  \caption{\label{tab:evaluation_tracks}Domain-specific evaluation tracks and their durations at the three main IR evaluation venues (CLEF, NTCIR, TREC). For each track, the track name and the domain are shown. Further columns represent years starting in 2000.}
  \includegraphics[width=\textwidth]{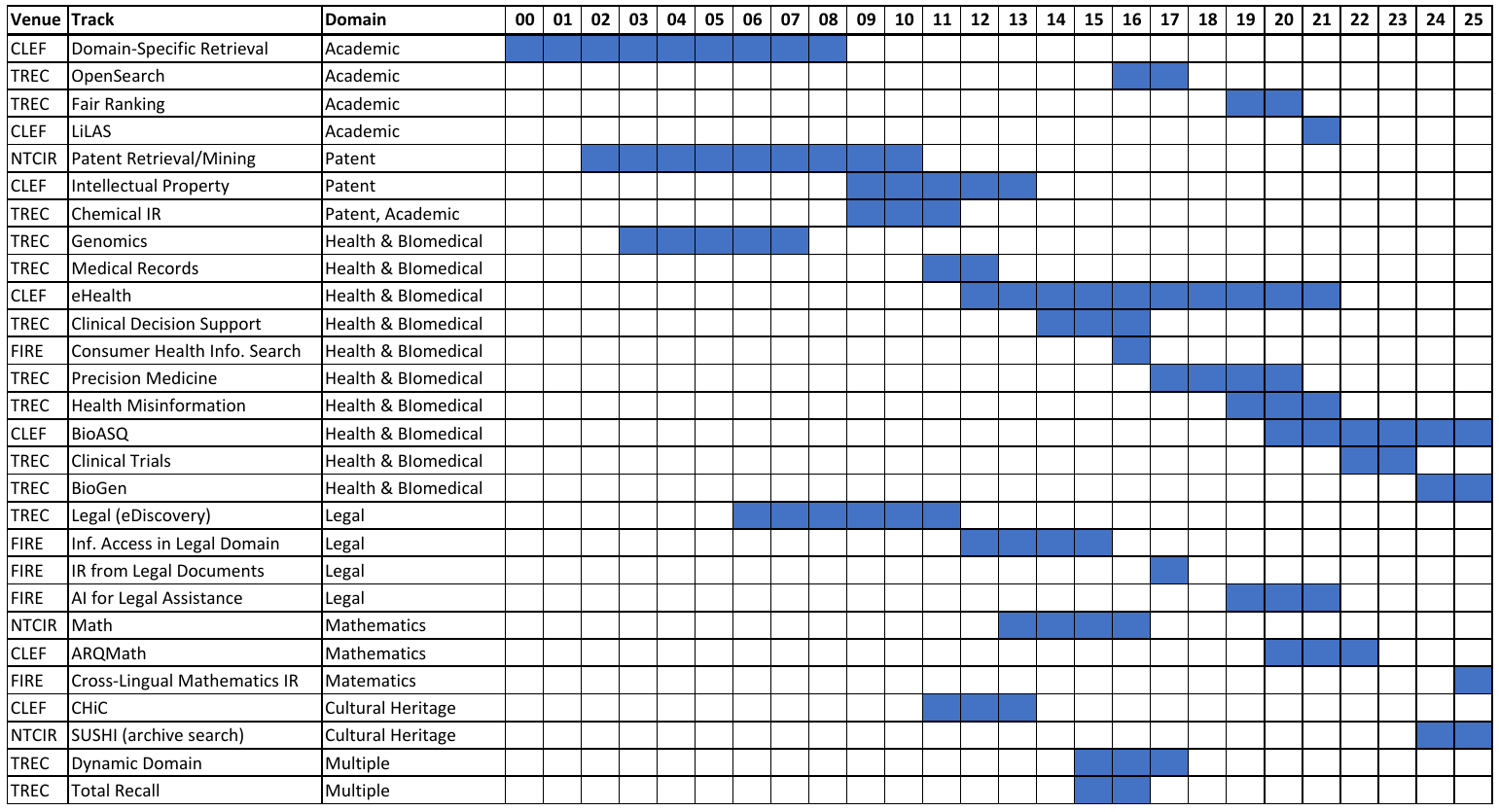}
\end{sidewaystable}

A further way to get an idea of the interest in the various domains over time is to follow the organisation of workshops in conjunction with IR conferences. Table~\ref{tab:workshops} shows the timeline of the workshops on domain-specific topics at the conferences SIGIR, ECIR and CIKM since the year 2000, structured in the same way as the previous table. 

\begin{sidewaystable}[htbp]
  \caption{\label{tab:workshops}Workshops related to domain specific IR and their durations at the SIGIR, ECIR and CIKM. For each workshop, the workshop name (occasionally truncated due to available space) and the domain are shown. Subsequent columns represent years starting in 2000.}
  \includegraphics[width=\textwidth]{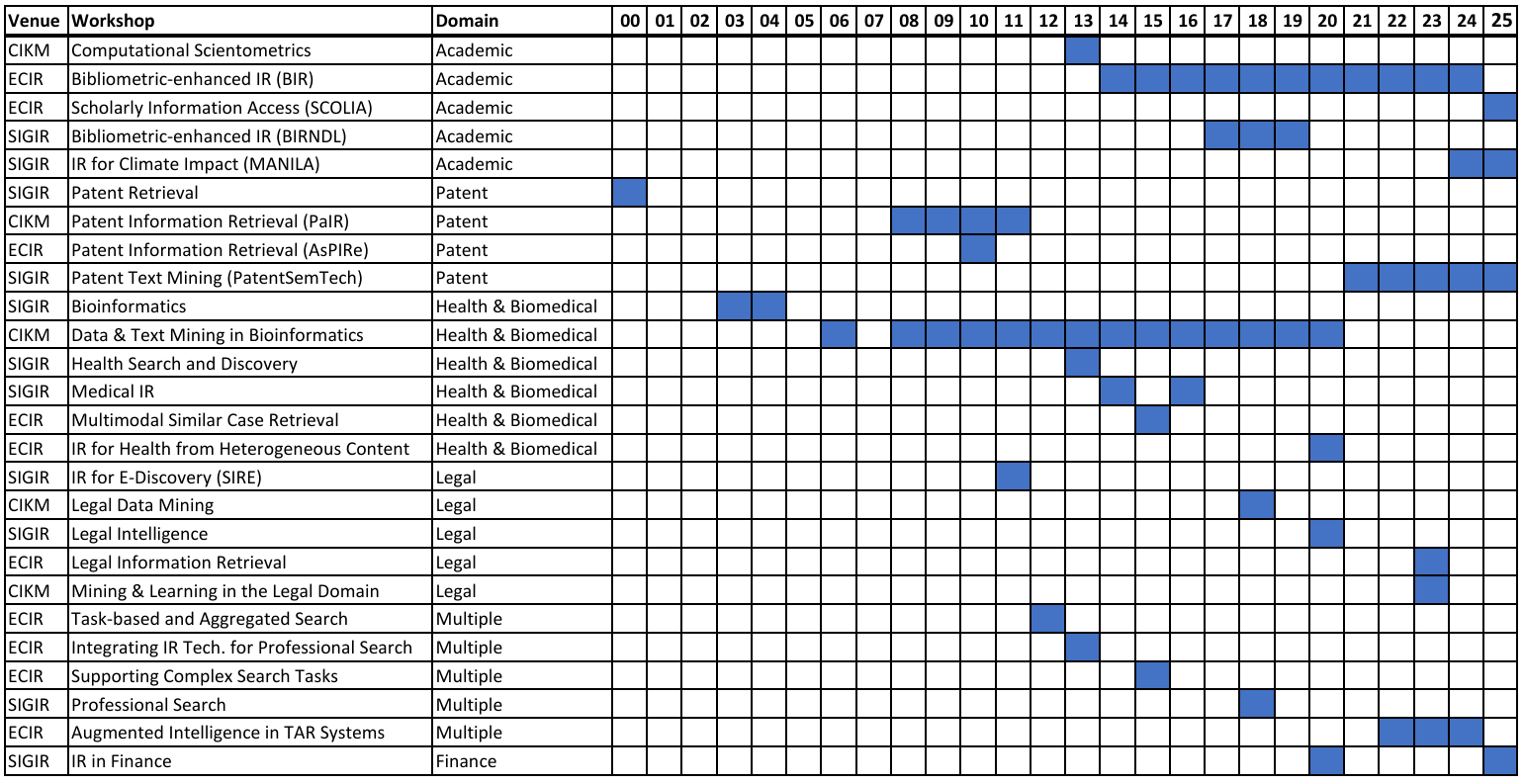}
\end{sidewaystable}

It is clear that there have been many workshops held only in a single year in the domains, with a few workshops held as series. The {\em Bibliometric-enhanced IR} (BIR) Workshop with the ECIR is the longest running workshop in the academic domain, running from 2014--2024, and now renamed to {\em Scholarly Information Access} (SCOLIA) from 2025. In the patent domain, the {\em Patent Information Retrieval} (PaIR) Workshop at the CIKM ran from 2008--2010 and the {\em Patent Text Mining} (PatentSemTech) Workshop, \revision{which started in 2019, and is since 2021 co-located with the SIGIR conference.} In the health and biomedical domain, the longest running workshop is the {\em Data and Text Mining in Bioinformatics} at the CIKM. The legal domain has not had a long running workshop in the IR community. A few workshops focussing on multiple domains had one year runs at the ECIR and SIGIR, while the {\em Augmented Intelligence in TAR systems} (ALTARS) Workshop focussing on Technology-Assisted Review in multiple domains (including the medical and legal domains) had the longest run from 2022 to 2024.

\section{Document Structure}
This chapter has introduced domain-specific search and the closely related topic of professional search and provided a framework for representing and discussing it. 
Chapter~\ref{chap:domain-specific_IR} takes a domain-centred approach and describes multiple domains in terms of the domain-specific search tasks and corresponding users and content. Recent IR work in each domain is covered, along with the availability of resources such as test collections. We then move to a domain-spanning approach in Chapter~\ref{chap:challenges} and detail multiple challenges that are each applicable to multiple domains. In Chapter~\ref{chap:retrieval_models_methods}, we describe models and methods recently developed in the IR community that overcome some of the challenges listed. The effect that Large Language Models are having on domain-specific search is discussed in Chapter~\ref{chap:LLMs}. Finally, the most promising directions for further development are presented in Chapter~\ref{chap:conclusions}.
\chapter{Domain-specific IR: tasks, users and content}
\label{chap:domain-specific_IR}
In this chapter, we discuss four domains that are prominent in IR research: legal, health and biomedical, academic, and patent. As introduced in Chapter~\ref{chap:intro}, these domains have large communities of professional searchers with standards for retrieval practices, but also common users conducting domain-specific web search tasks without adhering to particular standards. In other words: in each of the domains, a variety of tasks can be distinguished. For our four domains, we discuss the common tasks, challenges, and list the available benchmark collections.

\section{The legal domain}
\label{sec:legal}

Legal IR encompasses all search activities carried out in the legal domain\revision{, for work purposes,} by laywers, paralegals, legal scholars~\citep{makri_investigating_2008,van_opijnen_concept_2017,wiggers2023relevance}. The domain covers a range of tasks. The most prominent tasks in the context of IR benchmarking and evaluation are high-recall (or even: full-recall) tasks, i.e.
e-discovery~\citep{oard2013information,cormack2014evaluation} and case law retrieval~\citep{locke2022case}, but the area of legal information search is wider than these strictly defined tasks and includes also legal web search~\citep{wiggers2023relevance}, legal question answering~\citep{askari2023closer}, and legal news monitoring~\citep{sanchez2020easing}. 

The legal domain has clear national boundaries, with different law systems in different countries and documents published in local languages~\citep{wiggers2023relevance}. The national differences are also reflected in the benchmark datasets that are available, which focus on one language at the time (often English, but also Asian languages).

\paragraph{Case law retrieval} (also called \textit{legal case retrieval}) is the task of, given a new case, to identify prior cases that are relevant to it. This task is considered a high-recall or even a full-recall task that is carried out by lawyers and attorneys in common law systems. Similarly, statute or law retrieval is the task of, given a legal case, to identify which formal statutes and laws (regulations) are relevant to it. Both tasks are relevant preparations for a lawyer who represents a case. Interestingly, the text genres are quite divergent: where statutes and laws are written in very formal language, cases are generally more informal, describing the situation of real persons. 

Case law retrieval is considered a challenging task with notoriously low scores on benchmark data~\citep{goebel2023summary}. The first reason is that the documents in the collection are long and not all information in the document is relevant to the query. The second reason is that the query is a full document itself and again, not all information in it is relevant. Tasks that take a document as input for a search goal are query-by-document retrieval tasks: in case law retrieval, the input is a document describing a case. \revision{The challenges and characteristics of these tasks are detailed in Section~\ref{sec:qbd}.}

That case law retrieval is a difficult task, becomes immediately clear when we look at the results of \revision{the first task of the CoLIEE competition, the legal case retrieval task. In 2023 the highest F1 score was 0.3~\citep{goebel2023summary,li2023thuir}.}
%

\paragraph{E-Discovery} is the term for information retrieval tasks performed by legal professionals in the context of lawsuits and governance investigations~\citep{oard2010evaluation,grossman2011technology}. E-Discovery is considered a full-recall task, because any evidence that is missed by the professional might have legal implications. Because of the large collections and full-recall requirement, E-Discovery specialists are used to manually reviewing thousands of documents \revision{-- bills, e-mails, internal memos or documents, reports. Recently, they can make use of a set of automation techniques} called \textit{Technology Assisted Reviews} (TAR), which can be more effective than fully manually done reviews~\citep{grossman2011technology,cormack2016engineering,cormack2014evaluation}.

\paragraph{Legal web search} This form of legal Information Retrieval comes the closest to open\nobreakdash-domain web search: needs can be navigational and informational in nature (to speak in the taxonomy of~\cite{broder2002taxonomy}), and can be focussed on precision or recall or both~\citep{wiggers2023relevance}. A distinction from general web search is that the indices of legal web search engines are limited to field-specific documents, and that the access to those documents may depend on the user~\citep{wiggers2023bibliometric}. A similarity between open-domain web search and legal web search is that there is a large variety of document types \revision{to search through} and lengths of documents, ranging from government reports that can be hundreds of pages long to short (50-word) newspaper items. This distinguishes legal IR from patent search and academic search, where the collections usually contain documents of roughly similar length. Further, documents in a legal collection differ in terms of structure: While case law often follows a strict structure, like in patent search, other documents, like newspaper articles or legal blog posts, may have completely different formats. 

In legal web search, the aspects of documents that define relevance are partly overlapping with relevance aspects in general web search (e.g. title relevance, recency) but also include domain-specific aspects. A study among legal professionals and legal scholars indicated that the most important relevance factors for legal document retrieval are title relevance, document type, recency, level of depth, legal hierarchy, law area (topic), and authority/credibility~\citep{wiggers2022exploration}.

In the use of citations to other legal documents, the legal domain has similarities to the academic domain, although citations are sparser and do not occur among all types of legal documents~\citep{wiggers2022citation}. Where legal scholarly articles are similar to academic papers in other domains in terms of the use of citation references, more informal legal document types like legal blogs or news articles do not adhere to these strict structures.

\paragraph{Legal Question Answering}
Online communities \revision{are a space where} people can get answers to their (legal) questions. The information search behaviour in this context is completely different from the professional legal search tasks discussed above. An interesting aspect of legal QA is that the people asking the questions are typically laypeople (with very limited knowledge about the law and legal regulations), while people answering the questions are experts (legal professionals)~\citep{askari2023closer}. With the rise of generative large language models, there is research into the use of these models for answering legal questions. The results indicate that the best generative models (such as \revision{GPT-\revision{5.1}}\footnote{\revision{At the time of writing this document}}) can answer legal exam questions sufficiently~\citep{choi2023chatgpt,freitas2023does}, but this does not necessarily make them suitable as a consultancy tool for legal advice. An alternative, more trustworthy solution, is to use retrieval of expert answers from the collection of previous QA conversations on a large repository of legal advice~\citep{askari2023closer}. An alternative approach is using community QA data \revision{to find} legal experts~\citep{askari2022expert}. The use of retrieval models to retrieve expert answers as a safe and reliable alternative to answer generation with generative LLMs is further discussed in Section~\ref{sec:genLLMs}.

\paragraph{Legal test collections} A number of test collections have been developed for legal information retrieval, with a strong emphasis on high-recall tasks such as legal case retrieval. The TREC Total Recall Track~\citep{grossman2016trec} \revision{provided a frame for the researchers to explore technology-assisted review solutions, targeting exhaustive retrieval in large and homogeneous document sets.} The CoLIEE shared tasks~\citep{kano2018coliee,rabelo2020coliee} have focused on multiple subtasks including case law retrieval, legal question answering, and statute entailment, using data from the Canadian legal system. The FIRE AILA (AI for Legal Assistance) track~\citep{bhattacharya2019fire} addressed retrieval of prior cases and relevant statutes in Indian law, enabling research on multilingual and cross-genre legal texts. Datasets from non-English jurisdictions have also emerged, such as LeCaRD~\citep{ma2021lecard}, a Chinese legal case retrieval dataset, and CAIL2019\nobreakdash-SCM~\citep{xiao2019cail2019}, which supports similar case matching in Chinese civil law. While many of these datasets focus on case law, they vary in format and task framing. 


\section{Health and biomedical}\label{sec:health}
 The health and biomedical domain covers a wide range of tasks, including search in patient-specific documents (e.g. medical records) and search in non-patient-specific documents (e.g. medical publications). \citet{Her20} provides a general introduction to health and biomedical IR. One of the specifics of the health and biomedical domain are the extensive manually curated structured knowledge resources (e.g. Medical Subject Headings -- MeSH, International Classification of Diseases -- ICD), which can be used to enhance search as described in the survey by \citet{TG21}.
 
 A broad range of people carry out medical search, including citizens, research librarians, researchers, and clinical professionals. The tools in use range from web search engines to professional tools based on Boolean search. We concentrate in this section on three tasks that have been the focus of IR research in recent years.

\paragraph{Systematic Reviews}
\revision{As the basis for evidence\nobreakdash-based medicine, systematic reviews} involve the application of \revision{scientific information search methods}, in ways that limit bias, to the retrieval, critical appraisal, and synthesis of all relevant studies that address a specific clinical question \citep{CMH97}. Given a question about the efficacy of a certain medical intervention, the relevant primary studies (usually Clinical Trials) are retrieved, reviewed, and subjected to a statistical analysis (meta-analysis) to provide evidence in favour of or against the intervention. This eventually leads to the creation of clinical guidelines for medical practice. In order to limit the bias in systematic reviews, a standard protocol is followed \citep{TGC14} which includes the following tasks:
\begin{description}
    \item[preparation:] stating the research question and developing a search strategy\revision{ -- select the body of publications and reports to search into, specify a time frame, etc.}
    \item [retrieval:] performing a Boolean search on the literature, but also following references in retrieved papers to ensure that no relevant publication is missed. 
    \item [appraisal:] appraising the retrieved papers for relevance to the research question, also referred to as {\em citation screening}. This is usually done in two steps: in the first step only the title and abstract is considered while in the second step, the full text of the papers selected in the first step is considered.
    \item [synthesis:] Extract relevant information from the papers and perform a statistical analysis.
    \item [write-up:] The methodology for conducting the systematic review is written up and peer-reviewed before the systematic review is done. The final results are also written up, peer-reviewed and published.
\end{description}
Of most interest for IR are the retrieval and appraisal tasks, as reducing the number of irrelevant publications retrieved or using a classifier to carry out the appraisal task reduces the necessary human effort \citep{CHP06,BPT19}. \revision{Several types of approaches for these two tasks were evaluated, } including neural ranking \citep{WSK23}, random forest \citep{KEI15}, and Large Language Models~\citep{WSZ24}.

Nevertheless, not retrieving or automatically excluding a relevant publication can lead to a biased outcome of the Systematic Review as a whole. For this reason, specific evaluation measures have been developed for Systematic Reviews: One of the most commonly used is Work Saved over Sampling (WSS), introduced and described by \citet{CHP06} as
``the percentage of papers that meet the original search criteria that the reviewers do not have to read (because they have been screened out by the classifier)'' and further analysed by \citet{KLK23}. Later, \citet{KZK23} investigated how the influence of the inclusion or non-inclusion of a publication on the final outcome of the systematic review can be taken into account in the evaluation.

The demand for systematic reviews is increasing \citep{BGC10}, yet the process of creating systematic reviews is time-consuming, with an estimate of the mean time to complete and publish a systematic review of 67 weeks \citep{BBP17}. Furthermore, it is estimated that 23\% of published Systematic Reviews need updating within two years of completion \citep{SSA07}. Exceptional events such as pandemics also require rapid synthesis of evidence for effective interventions. Due to this, there have been efforts to increase the efficiency of performing systematic reviews, both on the methodological side through the development of a {\em Rapid Review} methodology \citep{GHT24} and the continually updated {\em Living Reviews} \citep{SES22}, as well as through developing technological support for the Systematic Review process \citep{BPE21}.   

Recent work in the IR community on search for systematic reviews in medicine includes evaluating the impact of query refinement \citep{SFC22}, and using Large Language Models to write complex Boolean queries \citep{wang2023can}.

\paragraph{Health Web Search}
Health web search is commonly associated with search for medical information by patients and individuals. However, a survey of 430 medical professionals (physicians in training, general practitioners, specialists, and professors) \citep{KGS13} revealed that general search engines are the most widely used source of medical information for the three categories excluding professors (for which it is the second most used source of information after medical research databases). Nevertheless, work in this area in the IR domain has focused on citizens and patients as the users of web search, although the assumption often made that patients know less than medical professionals doesn't always hold --- patients with chronic diseases may know more than some medical professionals about their diseases, as investigated in patients with type 1 diabetes by \citet{SHS13}. Extensive evaluation on consumer health search was done in the CLEF eHealth track \citep{SKG19} from 2013 to 2021.

\revision{To understand the information needs and search behaviour of different user groups, }\citet{PHM15} did an analysis of the search log files from three domain-specific medical search engines. Two of these search engines are used by medical professionals: TRIP\footnote{\url{https://www.tripdatabase.com/}} and ARRS Goldminer (no longer available) \citep{KC07}, while the Health on the Net (HON) search engine \citep{BGH17}, which allowed users to search through websites certified as trustworthy, and was targeted at laypeople (also no longer available\footnote{\revision{Health on the Net (HON) was discontinued in 2022. A snapshot can be found at \url{https://web.archive.org/web/20230709100236/http://www.hon.ch/en/}}}). An outcome of the analysis is that experts issue more queries and modify their queries more often, showing higher persistence than laypeople. The analysis also showed that diseases rather than symptoms were the focus of the largest number of sessions, indicating that the phenomenon of Cyberchondria -- ``unfounded escalation of concerns about common symptomatology, based on the review of search results and literature on the Web'' \citep{WH09} -- detected \revision{in the logs of the} general purpose web search engines is less prevalent on domain-specific search engines.

Given the wide range of knowledge \revision{possessed by the} people searching the Internet for health information, \citet{KJO17} investigated modelling users \revision{by means of two} cognitive models, representing users with, respectively, a low and high amount of medical and health-related knowledge. The cognitive models were built using Latent Semantic Analysis on a large
corpus of documents representative of the knowledge of the target user group. Additional support for users having lower medical and health-related knowledge has included: modifying the search engine ranking through the inclusion of text understandability as well as relevance information~\citep{PZH19}; bridging the gap between layperson and expert language by adding the most appropriate expert expression to queries submitted by users~\citep{SYY15}; \revision{linking informal medical terms in social media posts to their specialized counterparts and to Wikipedia explanatory articles \citep{NEPH24}}; and showing Health Cards, which are concise presentations of a health concept shown alongside search results for specific health queries~\citep{JZK19}.

A further important aspect of health web pages is their reliability or correctness. If people are influenced by incorrect information found in search results, they can make harmful decisions about the appropriate treatment \citep{PGS17}. This problem has been known since the early days of the web. An early solution was the manual certification of medical websites according to a set of principles~\citep{BGH17} -- this however did not prove sustainable, \revision{as seen in the case of the Health on the Net Foundation}. Automated approaches attempt to detect misinformation~\citep{ZZK22} or rank search results so that more reliable information is ranked higher~\citep{UV25}.

\paragraph{Medical Records and Clinical Trial Search}
One of the most promising IR tasks in this area is the retrieval of \revision{patients with similar medical issues~\citep{HLG21}}. If medical records of patients with diseases and disease \revision{histories similar to those of} a patient under consideration can be retrieved, then potentially promising treatment options could be identified.
However, research in the area of medical record search is complicated by the personal nature of the information in medical records and the difficulty of fully anonymising or de-identifying such information for data sharing. 
Hence the Information Retrieval tasks involving medical records at TREC have tended to use manually created synthetic cases. Because the creation of synthetic cases is time-consuming, these have rather been used as the queries rather than as the documents to search.\footnote{An exception is the TREC Clinical Decision Support Track 2016, which used the MIMIC-III dataset \citep{JPS16} of de-identified hospital admission data as queries} Hence the dominant paradigm has been linking patient-specific information to useful non-patient-specific information, such as the biomedical literature or clinical trials. The task of the TREC Clinical Decision Support Track (2014--2016) \citep{RDV16} was the retrieval of biomedical articles relevant for answering generic clinical questions about medical records. The TREC Precision Medicine Track (2017--2020) \citep{RDV20} had a similar biomedical article retrieval task but with a focus on cancer. The TREC Clinical Trials Track (2021--2023) focussed on matching patients to relevant clinical trials,\footnote{From \url{ClinicalTrials.gov}, a required registry for clinical trials in the United States.} in order to overcome the challenges of recruiting a sufficient number of patients for clinical trials \citep{RDV22}.

\paragraph{Health and biomedical test collections} Several test collections have been developed for health and biomedical IR, for different tasks and document types: The CLEF eHealth track~\citep{suominen2013overview,kelly2019overview} has supported a variety of shared tasks, including information extraction and retrieval from clinical narratives and consumer health documents. Systematic review search is targeted by the PICO~\citep{scells2017test} and CSMeD~\citep{KMS23} collections, both of which provide structured queries and annotated relevance judgments based on real-world clinical questions. The TripClick benchmark~\citep{RLS21,althammer2022tripjudge}, based on logs from a medical search engine, offers a large-scale, realistic evaluation setting for health-related web search. Several TREC tracks have addressed more specialized retrieval tasks: TREC Genomics (2003–2007)~\citep{HV08} explored retrieval of gene and protein-related information; TREC Medical Records (2011–2012)~\citep{Voo13} focused on patient-level retrieval from clinical records; and TREC Clinical Decision Support (2014–2016)~\citep{RDV16} evaluated ranking of biomedical literature for clinical care scenarios. Subsequent TREC initiatives further refined task granularity and content focus: the Precision Medicine track (2017–2020)~\citep{HRB20} targeted the retrieval of treatment-relevant literature for cancer patients, while the Clinical Trials track (2021–2023)~\citep{RDV22} addressed patient-to-trial matching, reflecting an increasing emphasis on personalization and real-world applicability in biomedical search.

\section{Academic} \label{sec:academic}
Academic search has played an important role in the history of the IR field. The field originated from the library and information sciences: in the time of paper archives, librarians were the first people who worked with (electronic) index files. In the 1970s scientific paper collections were the largest indexed file systems. The early studies of information seeking behaviour, from the \revision{first half of the} 1990s, also took place in the academic domain~\citep{kuhlthau_inside_1991,ellis_modeling_1993}. \revision{The study of academic search behaviour} in the age of large\nobreakdash-scale web search was later revisited by \cite{hemminger_information_2007}. 

An important characteristic of scientific articles is the use of citation references, connecting papers to each other \citep{cabanac2020scholarly}. We discuss this aspect in Section~\ref{sec:bibliometrics}~(\nameref{sec:bibliometrics}).

\paragraph{Academic search engines} Although the use of scientific papers as a means of publishing research results and the use of citations therein has not changed much over the past decades, the use of online resources for academic search has changed the way relevant papers are selected by scholars. \citet{gusenbauer2019google} provides an overview of online search engines and portals for scientific literature. Google Scholar, launched in 2004, has been the most used and biggest multi-disciplinary academic web search engine in the past two decades. Google Scholar was evaluated for systematic reviewing~\citep{giustini2013google,gusenbauer2020academic} and as a source of citation analysis~\citep{harzing2008google,martin2021google}, and conclusions have always been critical. Despite its limitations, Google Scholar has remained very popular, because of its easy-to-use interface and large scope of search results. Another successful and broadly used search engine has been Microsoft Academic (2016--2021)~\citep{wang2020microsoft}\footnote{Wikipedia has an extensive overview of academic search engines, both full-text aggregators and metadata services \url{https://en.wikipedia.org/wiki/List_of_academic_databases_and_search_engines}}. With the rise of retrieval-based chatbots like ChatGPT and Github Copilot, these interfaces have also gained use by academic searchers~\citep{liu2024chatgpt}.

\paragraph{Domains within academic IR}
The user-oriented research in the library and information sciences has traditionally focussed on the social sciences and humanities~\citep{kuhlthau_inside_1991,ellis_modeling_1993}. In 1980, Garfield pointed out the similarities and differences between information retrieval in the arts and humanities compared to the hard science \citep{garfield1980information}. This was followed-up in the 1990s and 2000s with studies investigating the scholarly behaviour of humanities scholars \citep{tibbo1992abstracting,cool1993information,buchanan2005information}. 

With the rise of the evaluation test collections (TREC) in the 1990s, the hard sciences and biomedical sciences became predominant in academic IR. 
In the hard sciences, literature reviewing is often not done with Boolean queries and explicit inclusion/exclusion criteria like in systematic reviews (see Section~\ref{sec:health}), but follows a more fuzzy approach with multiple different strategies. \citet{HagenBGKS16} show that for the task of finding related literature in computer science, a combination of multiple approaches is the most effective, combining citations, queries, and the `related articles' functionality of Google Scholar.

The humanities domain still receives attention from the library and information sciences field, because humanities scholars require access to archives and digital libraries to do their research. Applications in the humanities range from history~\citep{zhang2021review} and art history~\citep{verberne_information_2016}, to archaeology~\citep{brandsen2019user} and political studies~\citep{kaptein2010focused}. It was shown that in humanities research, the flows of data, tools, and tasks all need to be taken into consideration~\citep{koolen2020workflow}. It was also argued that information retrieval should be extended with text mining solutions in integrated working environments, as proposed by \citet{widlocher2015combining}.

\paragraph{Improving querying and ranking for academic search} Because it is common for information needs in academic search to require multiple queries and interaction with the search engine \revision{in order} to be satisfied, some work is directed at improved query formulation for academic search \citep{kim_automatic_2011,verberne2014query,li2017investigating}. Users can be helped with specifying their query through query term suggestion, although the effectiveness of this approach depends on the capacity of the user to select the best query term from the suggestions \citep{verberne_user_2015}. It was shown that personalisation hardly helps, but adding contextual information from the session can effectively improve query suggestions~\citep{li2017investigating}. Another direction for improving academic search is by connecting domain-specific queries and documents to explicit semantic information in the form of knowledge graphs~\citep{xiong2017explicit}. This is motivated by query log analysis revealing that it is challenging for search engines to identify the correct scientific concepts in search queries. On the collection side, retrieval can be improved by analyzing the internal structure of documents, e.g. by classifying the role of sentences in abstracts~\citep{hsu2021toward}, or by the use of citation clusters~\citep{bascur2023academic}. The latter study found that citation clusters can complement query-based search by finding additional relevant documents.

\paragraph{Academic search test collections} \revision{Presently, scientific papers can be collected in large numbers from preprint servers. Yet, even when capturing associated metadata (such as the The Semantic Scholar Open Research Corpus by~\citet{lo2020s2orc}), such accumulations of documents do not constitute proper test collections without queries and attached relevance assessments.}
Consequently, there are few test collections for \revision{academic search. Among them,} the iSearch collection of academic search tasks~\citep{lykke2010developing} is rich in terms of topic descriptions but shallow in terms of relevance assessments, and has only a small number of topics (see Section~\ref{sec:eval}). The SemanticSearch collection~\citep{xiong2017explicit} consists of 100 queries, a collection of scientific abstracts, and relevance assessments. The queries are extracted from the actual query log of the search engine, which makes them representative for real academic information needs.
\revision{Similarly, the LongEval Core collection~\citep{longeval_core_2025,longeval_clef_2025} makes use of search engine logs and click models to create relevance assessments for queries issued by users of the CORE system \citep{KHC23}.\footnote{CORE (COnnecting REpositories), \url{https://core.ac.uk/}}}  
The SciDocs collection~\citep{cohan2020specter} is a query-by-document collection (see Section~\ref{sec:qbd}) with titles and abstracts of papers. It does not have explicit relevance judgments, but instead uses statistics such as co-citation and co-read activities as similarity indicators. \revision{As a proxy for  relevance, c}itation metrics can be useful when evaluating text similarity, but excludes the evaluation of rankers that take bibliometric signals into account as this would entail a leak from the ground truth relevance to the ranking model. 

Despite these advancements, \citet{li2022academic} found that ``there is still a considerable gap between the search-needs of researchers and the usability of academic search engines. [...] Also, the evaluation, by academics, of academic search engines has not kept pace with the development, by search-engine providers, of those search engines.'' They provide a list of bugs encountered in academic search engines and recommendations for search engine developers.

\section{Patent}\label{sec:patents}

Conceptually, patents represent contracts between inventors and governments by which the former, in exchange for the disclosure of invention details, can exclude other parties from manufacturing and exploiting an invention without permission~\citep{piroi2019multilingual}. 

\begin{figure}[t]
    \centering
    \includegraphics[width=0.9\textwidth,alt=Patenting process - an overview of the main phases including search.]{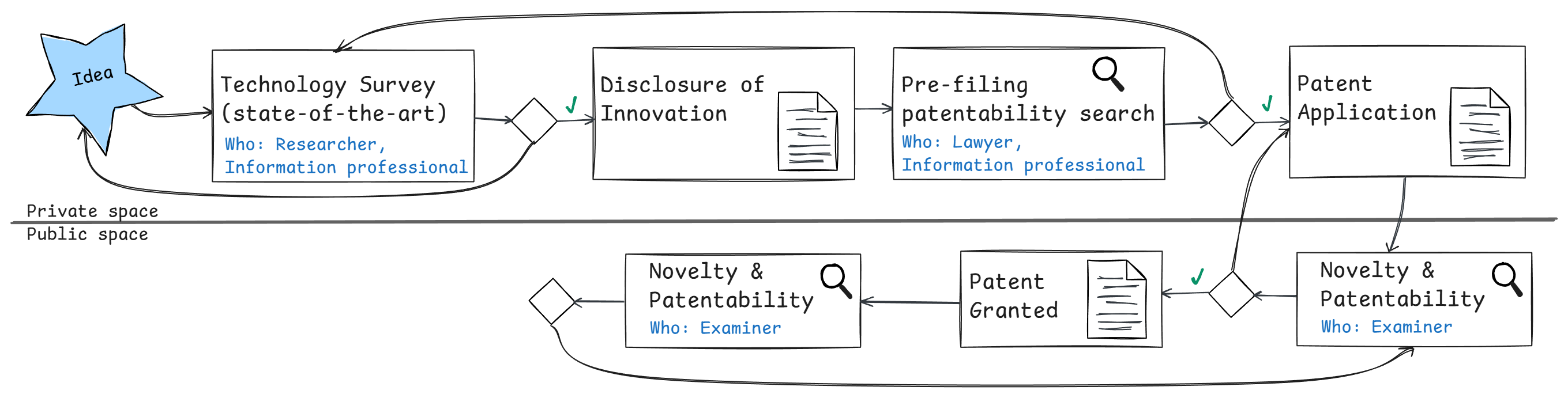}
    \caption{Main phases of the patenting process \citep[p15]{LH13}.}
    \label{fig:patenting_process}
\end{figure}

Turning an innovation idea into a granted patent involves several steps (Figure \ref{fig:patenting_process}). First, the inventor must conduct a technology survey before investing further R\&D resources. As the invention develops, a patentability search is done to find existing prior art pertaining to the core components of the invention. The inventor then files a patent application, which is reviewed by a patent examiner for novelty and patentability. During this process, there may be communication to revise or clarify the application. After the patent is granted, there is usually a 9-month period during which others can challenge its validity.

\paragraph{Patent search tasks} Patent search happens at different stages of the patenting lifecycle, and is performed by different stakeholders for different information needs. 
Due to the characteristics of the patent data, at a high level patent search can be also classified into \textbf{legal information search} and \textbf{technical information search}~\citep{schmoch_disclosure_1990}, which is not explicit in the patent search task taxonomy below. The former search is mostly done by attorneys who look at patent clauses, bibliographic metadata, patent jurisdictions, etc. The latter is mostly done by technical persons who look at the content of the patent documents, that is, the technical descriptions and drawings. Patent examiners at patent offices are trained to understand the claims' legal aspects as well as the technical content of a patent application document.
The major patent search types are (adapted from~\citet{alberts_introduction_2017}):

\begin{description}
    \item[State-of-the-art:] or \textbf{technology survey}, provides a comprehensive overview of a technology by examining all major aspects, using both patent and non-patent literature available up to the search date.
    \item [Prefiling patentability:] searches identifies prior art related to the invention’s core concepts and features, using all materials published before drafting the application.
    \item [Patentability:] or \textbf{novelty/prior art} searches identify published information that could prevent an invention from being patented, using all materials available up to the search or application filing date. The search is done, usually, by a patent examiner.
    \item [Freedom-to-operate:] or \textbf{infringement, right-to-use, clearance} searches identify enforceable patents or published applications (if no patent was granted) in a specific jurisdiction, selected based on where the product is planned to be made, used, or sold.
    \item [Validity:] or \textbf{invalidity, enforcement, readiness, or opposition} searches aim to find prior art published before the patent application filing date that could render a granted patent as unpatentable, using all relevant information available at that time.
    \item [Patent portfolio:] also called \textbf{due diligence} or \textbf{patent landscape} search, has a broad, business-driven information need to identify protection gaps, assess company value, or compare competitors. It uses worldwide patents, published applications, and business data.
\end{description}
In academic works \revision{on patent IR}, combinations of the tasks above are often wrongly melted into few information retrieval tasks, where the relevance relationship between a query (document) and the search result documents does not make explicit neither the many facets of a patent document, nor the particularities of the search task.

\paragraph{Patent test collections}What makes patent documents interesting for IR, is that to any one patent the patent offices publish more than one document. For example, for patent EP13165377\footnote{EP2660152 - Method for identifying an airplane in connection with parking of the airplane at a stand, \url{https://register.epo.org/application?number=EP13165377}} we identify, among others, three published documents: an application document without a search report, the search report, and the granted patent.
%
%
Viewing patents as the set of documents created during the patenting process (application, decisions, communication letters, search reports, etc.), the administrative aspect of the patent data becomes obvious~\citep{piroi2019multilingual}. The creation of the WIPO bibliographic standard\footnote{Standards (WIPO Handbook): \url{www.wipo.int/standards/en/}} is a direct consequence of this administrative side, where patent documents contain several data fields: bibliographic data (e.g. title, applicant names, addresses, filing dates), technical disclosure (e.g. abstract, descriptions, drawings), and legal information (e.g. claims, legal status).  Most patent offices that publish patent documents adhere to this standard.

The main source of patent data are the national and international patent offices. WIPO maintains a Patent Register Portal\footnote{Patent Register Portal. \url{https://inspire.wipo.int/patent-register-portal}} which points to on-line patent office databases around the world. For the purpose of IR experiments in the patent domain, the need for appropriate test collections has been made explicit in 2000~\citep{kando_what_2000}. Soon after, NTCIR organized the first retrieval evaluation campaign using patent data. 
In 2009, further patent test collections were created, under the umbrella of CLEF and TREC. 
The CLEF-IP test collection contains  patent documents issued by the European Patent Office (EPO)~\citep{clef_2013_dataset_2021,clef_2009_dataset_2021,clef_2010_dataset_2021}, while the TREC-Chem collection contains patents in the chemical domain published by EPO and the US Patent Office~\citep{lupu_overview_2009}.

More patent text collections have been released in the last decade, however, compared to CLEF-IP and TREC-Chem, they don't contain ground truth to assess the efficiency of models on a given retrieval tasks. These are collections of patent texts, sometimes including images as well, which researchers can use for patent data mining and foster good scientific practice by using them to compare models. 
Such examples include the WPI test collection \revision{and the WPI+ test collection~\citep{lupu_2019_WPI,salampasis_et_al_2025}}, a multilingual collection of patents obtained from major patent offices (USPTO, EPO, CN, JP), the Linked Open EP Data~\citep{LinkedEPData} of URIs to EP patent applications, with documents in either German, English, or French. The EPO makes available also full-text data in XML and PDF formats~\citep{EPFulltext}. Since the availability of Google Patents\footnote{Google Patents, \url{https://patents.google.com/}, accessed June 2025.} this dedicated search engine has been used to create ad-hoc collections for particular research questions (e.g., linking engineering design stages with knowledge disclosed in patents~\citep{Jiang03062023}, patent summarization~\citep{Girthana_pat_summarization_2019, casola_2022}, or classification~\citep{li-etal-2024-scalable}).


\section{Cross-domain benchmark data }
\label{sec:crossdomain}
In a broad sense, cross-domain research (be it multi\nobreakdash-source and multi\nobreakdash-domain data mining \citep{2015cross_domain_big_data} or recommender systems \citep{2022cross_domain_rec}) refers to applying methods developed in one domain (e.g., health, patents) to other domains (legal or scientific). 
%
%
For example, the BERT\nobreakdash-PLI model~\citep{shao2020bert} for paragraph retrieval in the legal domain has been tested on patent retrieval by~\citet{althammer_cross-domain_2021}.
Domains are often not formally defined, the common characteristics of documents making them belong to one domain being kept at a high\nobreakdash-level. For example, the patent domain contains documents pertaining to patents, the health domain may contain medical scientific publications but also electronic health records, while the academic domain looks at scientific publications in general, without necessarily considering different scientific areas or subject classifications.

In the IR community, cross-domain retrieval is often understood to be multi\nobreakdash-domain or multi\nobreakdash-purpose retrieval, that is, retrieval from data that covers multiple domains. Models using cross-domain data sets range from federated retrieval~\citep{milad_shokouhi_federated} to question answering and learning-to-rank~\citep{chen_knowledge_2010}. Cross-domain benchmark datasets are often a compilation of other datasets and collections, making them diverse in content and especially appropriate to training neural network-based models. One such dataset is BEIR~\citep{thakur2021beir} which is comprised of 18 retrieval datasets from domains like News, Academic publications, Wikipedia, Bio-Medical, social media posts (Twitter), and financial data. The breadth of IR tasks that BEIR covers range from bio-medical IR to fact checking and entity retrieval.

The availability of cross-domain benchmarks for the design and training of IR systems is significant to their robustness and generalizability. A source of creating such benchmarks that is in available reach to researchers is the World Wide Web. Collaborations between academia and companies operating search engines have resulted in the publication of text collections for training IR models. MSMARCO~\citep{Bajaj2016MSMA}, LETOR~\citep{qin2010letor}, ClueWeb22~\citep{overwijk_clueweb22} (all resulting from collaborations with Microsoft), LongEval Retrieval Collection~\citep{galuscakova_longeval-retrieval_2023,2025_longevalweb} (from collaborations with Qwant), or LongEval CORE collections~\citep{LongEvalCore_train,LongEvalCore_test} for scientific retrieval (from collaboration with CORE--COnnecting REpositories) are such examples of cross-domain benchmarks.
%

Recently, retrieval based question answering tasks have developed their own cross-domain benchmarks. One such example is the MultiReQA collection~\citep{guo_multireqa_2021}, which combines  a number of datasets developed for the MRQA Shared Task~\citep{fisch-etal-2019-mrqa}. 

When ranked lists of results are of importance, learning to rank is an IR concept employed in many retrieval model architectures. The cross domain learning to rank problem has been formally defined by \citep{chen_knowledge_2010}, who, in their work, made use of the LETOR benchmark~\citep{qin2010letor}, a collection of documents first published in 2007, combining data from TREC 2003 and 2004 web tracks and a subset of the MEDLINE database of medical publications.


\chapter{Challenges of domain-specific IR}
\label{chap:challenges}

We outline the challenges of domain-specific IR: long documents and long queries, complex and specific user needs, session-based and exploratory search, reproducibility, bias and fairness, and evaluation. Not all these challenges are unique to domain-specific applications; especially the latter three are relevant aspects of many or even all IR problems. We discuss them in the light of our domains, and highlight literature that discuss these challenges in domain-specific contexts.
\section{Long documents and long queries} \label{sec:long}

In domain-specific IR, we often deal with relatively long documents. Examples are scientific articles and patent documents (see Chapter~\ref{chap:domain-specific_IR}). Ranking these documents poses challenges for models because relevant information might be dispersed throughout the document~\citep{callan1994passage}. Notably, the complete document is not necessarily relevant to the user, but only parts of it. Apart from these fundamental challenges of information in long documents, long documents may also pose practical and technical challenges to retrieval models. 

\paragraph{Challenges of long documents for retrieval models} If we organize ranking models in terms of the representations of document content and queries, there are two types of ranking models: lexical rankers (sparse models) and embeddings-based rankers (dense models). Lexical rankers such as BM25~\citep{robertson1994some,robertson2009probabilistic} represent documents and queries by the words they contain, weighted by frequency. Embedding-based rankers encode documents as dense embeddings, typically created with transformer encoders (BERT models). With lexical rankers, a document can only be retrieved for a query if there is word overlap between the two. An embeddings-based ranker might retrieve documents that do not contain query terms, but semantically similar terms. In research and practice, both rankers are commonly used and often combined~\citep{lin2022pretrained}. 

BERT-based rankers are technically limited because of the computational complexity of their text processing approach \citep{nogueira2019passage,yan2019idst}: a transformer encoder computes the attention weight (dot product) between each pair of input tokens. When encoding a document, this leads to quadratic complexity relative to the length of the document, therefore posing a high load on the computational memory. For this reason, BERT models have been by default limited to an input length of 512 tokens. Transformer architectures that can cover longer input sequences such as Longformer \citep{beltagy2020longformer} have not shown high effectiveness in retrieval tasks, perhaps due to their sparsified attention mechanism: \revision{the model combines local attention using a sliding window with global attention on a few selected tokens. This reduces the memory burden of the attention mechanism, but also limits the context that individual tokens are compared to.}

\revision{In contrast, lexical models are technically capable of processing long documents, with some lexical rankers including a document length weighting mechanism. The most prominent example is BM25 where the document length is accounted} for in the estimation of relevance~\citep{robertson1994some}. 

\paragraph{Document lengths in benchmark data} The most commonly used IR benchmark is the MS Marco passage dataset~\citep{Bajaj2016MSMA}. The median length of documents (passages) in this benchmark is 50 words. This is convenient for the evaluation of BERT-based rankers, which by default are limited in the length of the input that they get, but it is not a very realistic evaluation setting, especially not in domain-specific IR. In the heterogeneous BEIR benchmark~\citep{thakur2021beir}, the datasets with the longest documents are TREC-News (635 words on average) and Robust04 (466 words on average). Still, this is relatively short compared to patents: in the CLEF-IP 2011 collection the average document length is 5774 words~\citep{mahdabi2011report}.
In Section~\ref{sec:dealingwithlong} we discuss solutions to dealing with long documents, such as paragraph-level retrieval and aggregation.

\paragraph{Query-by-document retrieval}\label{sec:qbd}
In some domain-specific tasks, queries can become extremely long. Apart from Boolean queries -- which we will discuss in the next section -- this is particularly the case in tasks where documents are used as queries, so-called `query-by-document retrieval' tasks or `query-by-example' tasks \citep{yang2009query}. Key examples of such tasks are legal case retrieval \citep{rabelo2020coliee}, patent prior art retrieval \citep{piroi2011clef,piroi2019multilingual}, and finding similar articles in scientific literature retrieval \citep{cohan2020specter}. In the previous section, we have discussed two challenges of long documents in the collection: conceptually (information spread over multiple parts of the document) and computationally (long input processing for transformer models). When dealing with documents as queries, the challenge is even bigger, because the \textit{information need} is spread over a long text, \revision{and shows a composite characteristic, with various parts of the text contributing to parts of the overall information need (see next Section)}. So the main challenge is to process the whole query, both computationally and in terms of the user need.

\section{Complex and specific user needs} \label{sec:complex_needs}
Information needs are often more complex and more specific in domain\nobreakdash-specific search than in open\nobreakdash-domain web search. 

\paragraph{Complexity} of search tasks \revision{includes three dimensions \citep{wildemuth2014untangling}, which we detail in the following: (1) multiplicity of subtasks or steps, (2) multiplicity of facets, and (3) indeterminability.}

\revision{(1) Number of subtasks: complex search tasks may need to be split into a number of \textbf{subtasks} \citep{madden2006search,albertson2011situated,kulahcioglu_incorporating_2017}, often by collecting information from multiple sources \citep{elsweiler2007towards, he_characterizing_2013}. The collection described in~\citet{verberne2019information} contains a set of search tasks, freely described by the professional searchers who worked on these information seeking problems. An example of a complex information need in that dataset is ``evidence of effectiveness of rotational atherectomy in the treatment of calcified coronaries''. When asked to describe which actions the professional took to search, they replied: ``Medline, Google Scholar, Trust's discovery system. Searched for combinations of: calcified coronaries; coronary artery calcification CAC; calcified lesion; undilatable lesion; rotational atherectomy; rotablation; orbital atherectomy. Filtered this to last 10 years.''}

\revision{(2)  Multiple facets: complex search tasks may be composed of multiple facets (\textbf{subtopics}). This has been operationalized in the pre\nobreakdash-web search literature by the number of components in Boolean queries \citep{williams1986comparative}, or by the number of concepts included in the search \citep{iivonen1995consistency}. \citet{capra2007effects} differentiates between simple tasks consisting of just one facet and complex tasks consisting of multiple facets. }

\revision{(3) Complex search tasks often are \textbf{indeterminate}, uncertain, or ill-defined~\citep{singer2012ordinary} with respect to the task process or goals. This indeterminate nature is an additional reason for a search task to be split in sub-tasks \citep{bystrom1995task}. Indeterminability of a search task is also related to exploratory search (see Section~\ref{sec:exploratory}).}
\paragraph{Specificity} of information needs \revision{can be observed through three aspects: (1) context\nobreakdash-specificity, (2) user\nobreakdash-specificity, and (3) topic\nobreakdash-specificity.}

\revision{(1) With respect to \textbf{context\nobreakdash-specificity}, }information needs have a surrounding seeking context and a task context, as illustrated in the model of task-based information seeking by \citet{ingwersen_turn:_2005} (Figure~\ref{fig:ingwersencontext}). This context determines the meaning of the issued queries. Because of the importance of context, a single information need might have a different meaning in one domain than in another, and the context co-defines the relevance. For example, an information retrieval researcher might issue the query `search behaviour', aiming to find literature on the behaviour of users of search engines in their search for information. A marine biologist might issue the same query `search behaviour', aiming to find literature on the behaviour of sea mammals searching for prey. 

(2) \textbf{User-specificity} \revision{of the information need refers to how individual users,} even within one same domain, work on their own specific topic and have their individual information need. From the search engine perspective, two users who issue the same query may both have a different underlying information need. A study analysing the queries issued in a legal web search engine indicated that the majority of queries issued by one user in the time span of a month is unique to that user~\citep{wiggers2024high}. This is the consequence of the legal search engine users being experts that have highly specific information needs. 

(3) \revision{Finally, by the \textbf{topic-specificity} we understand that }the topic of a query in domain-specific tasks can be infrequent and unique.

\revision{The distribution of the search needs in domain\nobreakdash-specific search has a long tail that translates not only into} challenges for the users but also for the search engines~\citep{goel2010anatomy}: The specificity of the information need makes the search space of relevant information in the collection small (less redundant) and, at the same time, makes it more challenging to identify the relevant information.  
\paragraph{High-recall information needs} One particular type of specific user needs are high-recall search tasks, which occur in the legal and patent domains (see Sections~\ref{sec:legal} and~\ref{sec:patents}), but also in systematic reviews in the biomedicine domain (see Section~\ref{sec:health}). In high-recall tasks, the use of Boolean queries is common, giving users full control over the result set, yet they are challenging to construct~\citep{karimi2010boolean,kim_automatic_2011,verberne2016boolean}. Further, there is a gap between the evaluation of relevance ranking in IR contexts, where there are limited relevance assessments for a set of user queries (described by a short topic summary and a keyword query) and the practice of Boolean query for full control in high-recall tasks. See Section~\ref{sec:queryproc} for a description of methods for processing Boolean queries.

\section{Session-based and exploratory search} \label{sec:exploratory}

It is common for professional information needs to require multiple queries \revision{in order for the need} to be satisfied. A sequence of queries pertaining to one information need can be referred to as a \textit{search session}. Pragmatically, however, in the context of search log analysis, search sessions are commonly defined as periods of 30 minutes of interaction between a user and a search engine, irrespective of the topics being searched. \citet{wiggers_task_2022} show that, in the context of legal web search, the 30\nobreakdash-minute heuristic is not suitable for accurately grouping related search queries from a single user into a specific search task. Through manual annotation of the search engine's query log, 
the authors identify 55 individual search tasks \revision{spread over 126 sessions of 30 minutes. This indicates that, in professional search,} one task can span multiple sessions.

When studying session-based search, the success of the search is not evaluated by an individual query result, but after the completion of the whole session~\citep{ruotsalo_interactive_2020}. This also relates to models of information seeking in context, such as depicted in Figure~\ref{fig:ingwersencontext}: for information needs that require multiple queries, it is the task success that matters and not the individual query result evaluation. Although the IR field acknowledges the importance of session-based evaluation~\citep{belkin2012whole,wicaksono_modeling_2021}, most experimental evaluation takes place with test collections at the query level~\citep{lin2022pretrained}. An exception is the TREC Session Track~\citep{carterette2014overview}, which ran from 2011 to 2014. The goal of the track was to analyse if incorporating previous queries, ranked results, and user interactions could enhance the effectiveness of a query's retrieval. We note that the collections used in the track are web-based and not domain-specific.

\begin{figure}[t]
    \centering
    \includegraphics[width=0.8\linewidth]{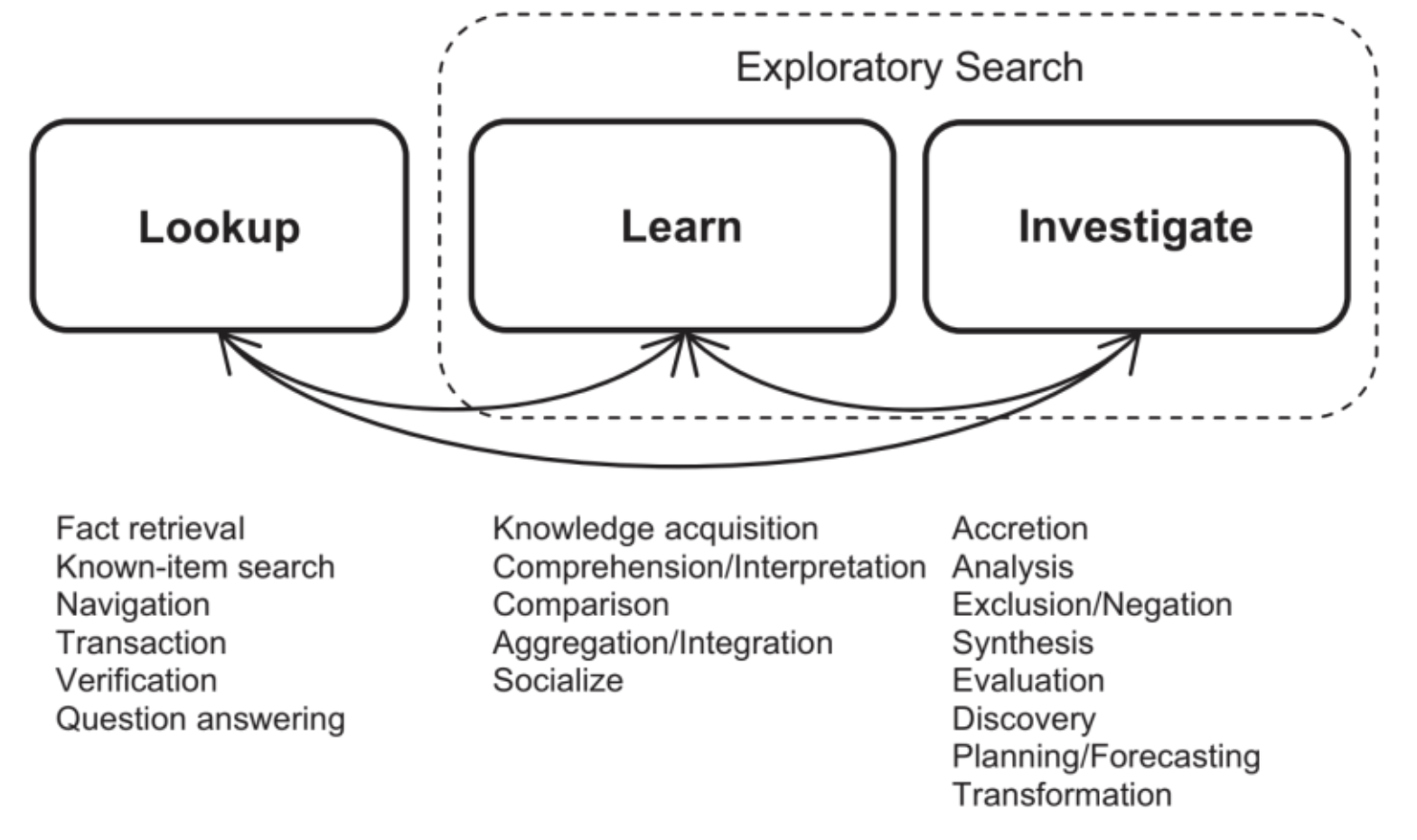}
    \caption{\revision{Definition of exploratory search activities by~\citet{white2009exploratory}, based on \citet{marchionini_exploratory_2006}.}}
    \label{fig:exploratorysearch}
\end{figure}

In the context of domain-specific search, session-based evaluation is even more important than in web search, because of the typical search needs of professionals. Not only are domain-specific information needs often more complex and specific (as described in Section~\ref{sec:complex_needs}), but it is also assumed that professionals have \textbf{exploratory search needs}~\citep{he_characterizing_2013}. 

\revision{\citet{marchionini_exploratory_2006} defines exploratory search tasks as tasks that require higher-order intellectual activities. \citet{wildemuth2012assigning} describe exploratory search needs as non-specific, open-ended, and uncertain, involving multiple sources, and ``being motivated by ill-defined or ill-structured problems; they are
dynamic and evolve over time; they are multi-faceted and may be procedurally complex; and they are often accompanied by other information or cognitive behaviours, such as sensemaking.'' \citet{white2009exploratory} distinguish between lookup searches and exploratory searches in terms of the \textit{activities} that are needed to complete the search tasks: ``activities associated with exploratory searches require more involvement from the user, more synergy between user and search system, and more functionality from the system, extending beyond just query specification and result presentation'' \revision{(see also Figure \ref{fig:exploratorysearch})}. In other words, exploratory information needs are information needs that require multiple queries and include browsing and analysing multiple documents in order to be satisfied~\citep{makri_investigating_2008,white2009exploratory}. } 

Exploratory search occurs commonly in situations where the searcher does not know yet how to exactly specify the requested information, and uses the search process to identify the relevance aspects of the search topic. This is also common in educational settings~\citep{ghosh_session_based_2019}. A qualitative study of a small collection of professional search needs, published by~\citet{verberne2019information}, indicates that multiple sources are used in many of the search tasks. This is probably caused partly by the importance of completeness of the result set (systematic review tasks, which are inherently recall-oriented). It is not straightforward to identify the exploratory nature of search tasks in an offline collection, where most of the topics seem well-defined and specific rather than exploratory.

\section{Reproducibility} 
\label{sec:transparency_reproducibility}
In IR as a scientific field~\citep{luk_why_2022}, evaluation of IR systems via test collections is, by design, oriented towards reproducibility. Despite the existence of reusable test collections, issues with reproducibility may arise due to differences in preprocessing, different implementations of the same ranking model~\citep{kamphuis2020bm25}, and different versions of evaluation metrics.  Attempts to replicate or reproduce previously published scientific results rightfully receives attention in the scientific IR community~\citep{ferro_introduction_2018,breuer2020measure,lin2020reproducibility,keller2024replicability}. Reproducibility of domain-specific IR poses additional challenges because of the relative small test collection size~\citep{althammer_cross-domain_2021}\revision{, which makes differentiating between systems difficult}, and the importance of user-centric online evaluation \citep{breuer_reproducible_2020}. 

Besides the importance of reproducibility of \textit{IR research outcomes} for IR researchers, reproducibility of \textit{search strategies} is important for professional searchers in their information seeking process, especially in the context of systematic reviewing. Reproducibility is not equally important in all domain-specific search tasks; the emphasis on reproducibility mainly relates to highly controlled professional search tasks where the set of retrieved results has implications. Examples are patent prior art search and systematic reviews~\citep{rethlefsen2024systematic,koffel2016reproducibility}. 
Reproducibility of search strategies refers to three notions: (1) when the same user enters the same query in the same search engine twice, at different times, they expect the same search results; (2) when a searcher informs a collaborator which query terms will produce certain requested search results, they should be able to trust the search engine to produce and show the same results to the other searcher; (3) in the context of systematic reviewing or prior art search, the published search strategies and queries should lead to a fixed set of search results. 
A 2024 study investigated how many database searches are reproducible, measured by retrieving the same number of results as the original study (with a 10\% difference allowed). The finding was that only 5\% of database searches was reproducible according to this definition and only one systematic review article provided all the necessary search details to be fully reproducible \citep{rethlefsen2024systematic}. \citet{SPR24} present a retrieval system that maintains a versioned and time-stamped index. The time-stamps allow re-execution of previously posed queries resulting in the same ranked list. 


\section{Bias and fairness}
\label{sec:bias_fairness}
In IR, fairness is commonly defined as lack of bias; fairness and bias can be considered two sides of the same coin~\citep{gao2019fair}. In the search engine context, we can distinguish multiple types of bias\revision{, not mutually exclusive, namely exposure bias, societal bias, and interaction bias.}


\textbf{Exposure bias}\revision{, related to what \citet{baeza2018bias} calls \textit{activity bias}, refers to situations where there is a large amount of web content that is never visited by anyone. For example, 31\% of the Wikipedia articles added or edited in May 2014 were never visited in the next month. In connection to search engines,} a document from a popular web domain (e.g. Wikipedia, Stackoverflow) may be ranked higher than a document from a non-popular source, despite its topical relevance. This is related to \textit{retrievability}~\citep{azzopardi2008retrievability}, that is, does every document have an equally fair chance of being retrieved, or is the search system biased towards some documents over others. In other words, is the information exposure in the search engine fair~\citep{gao_fair_2022,}? This is a form of bias that is particularly relevant in domain\nobreakdash-specific contexts. For example, we typically want the retrievability of relevant scientific or biomedical documents to be fair with respect to meta-data aspects such as authorship or publication venue. \citet{mcdonald2022search} investigated exposure bias mitigation through diversification in academic search. Using data from the TREC Fair ranking track, they found that explicit search results diversification performed better than implicit diversification for providing fair exposure of authors from disadvantaged groups. 

\textbf{Societal bias }\revision{ can be caused by \textit{data bias} or \textit{algorithmic bias} \citep{baeza2018bias}. Data bias originates from the underlying collection, e.g. biographies of females on Wikipedia being under\nobreakdash-represented, or a much higher frequency of males than females for the topic `painters from the 17th century'. Algorithmic bias originates from the text representation model (the pre-trained BERT embeddings underlying the supervised ranker), or by the rank training that prefers some documents over others, e.g. a search engine showing results for male football players and not of female football players when the query is gender neutral, `football players'}~\citep{bigdeli2022gender,abolghasemi2024measuring}. These aspects, however, are difficult to disentangle~\citep{abolghasemi2024measuring}. In relatively technical domains like medicine, law, and science, the presence of societal bias in search results is less obvious than in open-domain web search.  
    
\revision{\textbf{Interaction bias} originates from the \textit{user interface} or the user's self-selected \textit{biased interaction} \citep{baeza2018bias}, e.g. confirmation bias in health search \citep{mansour2023measurement}}. In search and recommendation systems with ranked results presentation, a user is more likely to click on a document that is positioned high in the result ranking or prominent on the result page than a document that is hidden at a lower rank or even on the next page (\textit{position bias})~\citep{joachims2017unbiased,wang2018position}. This is a form of bias that is inevitable because every search engine interface shows some documents (top-ranked) more prominently than other (lower-ranked). In high-recall tasks, the effect of position bias is smaller because the user is motivated to look beyond the first result page to lower-ranked documents. In other words: the more ad-hoc the search is, the stronger the bias caused by the user interface; a more persevering (professional) user is less affected by the presentation of the documents than an ad-hoc searcher \citep{verberne_user_2015}. \revision{In health information seeking, which is largely conducted by non-experts, content biases combined with interaction biases, can lead to incorrect conclusions \citep{white2014content,hashavit2021understanding}}

\revision{From the academic perspective, we add one additional form of bias:} \textbf{Bias in evaluation}. Test collections consist of a document collection, a set of queries, and a set of relevance assessments for a subset of query\nobreakdash-document pairs. \revision{The set of relevance assessments is customarily created by evaluating documents pooled from the list of results submitted by participants to an shared task.} When pooling the documents for relevance assessment, a challenge is that the set will be biased towards the rankers that were used to create the pool (\textit{pool bias})~\citep{lipani2016fairness}. This problem is prominent in domain-specific collections because relevance assessments by domain experts are expensive to create so it is of importance to create a good pool of documents for hand\nobreakdash-labeling~\citep{althammer2022tripjudge,lipani_fixed_cost_2017}. 

When optimizing for fairness, the developer has the choice to prioritize parity-based group fairness or opt for disparate treatment fairness \citep{gao_fair_2022}. In the first case, every group or topic has equal opportunity to be retrieved. In static ranking, this means equal numbers of each group should be presented in the ranking (e.g., equal number of documents giving benefits and risks of coffee; equal number of male and female football players). In the second case, the proportion of items from each group should be proportional to the average relevance of each group (e.g. more documents describing benefits than risks of vaccinations, in line with the most trustworthy sources).




\section{Evaluation of domain-specific IR} \label{sec:eval}

\paragraph{Challenges} There are a few characteristics of domain-specific IR that make evaluation challenging~\citep{wiggers2024high}. First, explicit relevance judgements are expensive to collect because most users of domain-specific search systems are highly skilled professionals (e.g. lawyers, scholars, medical doctors). Second, in professional search tasks, high recall is often necessary because the cost of missing results is high (e.g. in the patent domain, legal IR, or systematic reviewing), while at the same time, precision at high ranks is required because professionals work under time pressure and therefore are not very tolerant to browsing many irrelevant results. Third, since information needs in domain-specific search are highly user-specific, the number of user interactions that can be used for implicit feedback is limited. Fourth, in many professional IR systems, different users see different results in their results list, based on the journals/sources they are subscribed to, and thus have access to. Fifth, recency is considered very important, playing a large role in the ranking algorithm. Because of this, and the high frequency with which new documents are published, the top of the results list is highly dynamic \revision{(an illustration of this fact is presented by~\citet{staudinger_timir_2025})}.

It is relevant to evaluate the effectiveness of IR models in the context of users' real search tasks. Ideally, this is measured in a live, online search engine, but this is challenging because of the sparseness and uniqueness of user queries. In some evaluation studies, log data in a live search engine is used, although suffering from sparseness \citep{wiggers2023bibliometric}. User observation studies are typically small-scale but can provide valuable information about the usability of the search engine~\citep{verberne_information_2016,brandsen2021usability}. In the remainder of this section, we discuss two most common forms of evaluation in domain-specific search: benchmarking with static test collections and simulation studies.

\paragraph{Static test collections} Evaluation in IR mainly relies on static test collections, consisting of a document collection, queries, and relevance assessments for query-document pairs. These test collections have been created in successful benchmarking initiatives in TREC~\citep{voorhees2000report} and CLEF~\citep{goeuriot2017clef,piroi2013clef}. For domain\nobreakdash-specific tasks that often take place in professional contexts, creating relevance assessments can be an expensive undertaking. As mentioned above, relevance assessments by experts are expensive. As a result, the existing benchmarking collections often suffer from shallow relevance assessments~\citep{althammer2022tripjudge}: each query only has a small number of assessed documents, and hence, a small number of relevant documents.  Unassessed documents will all be marked as non-relevant in the evaluation~\citep{webber2009score}. This hampers evaluation because the comparison of ranking models is not sufficiently fine-grained if many documents are unassessed. This is an even larger problem when we deal with high-recall tasks, where we need deep relevance assessments but an insufficient, shallow pool with only a few assessed documents per query is available. Take, for example, the iSearch collection, which is a small collection of scientific search tasks in the physics domain~\citep{lykke2010developing}. The collection has a rich definition of search tasks, including explicit information needs, ideal answers, and search terms. The relevance assessments, however, are shallow and relevant documents are unevenly distributed over topics, which makes benchmarking on this data problematic \citep{CarevicS14,verberne_author-topic_2018}. 

Some domains can rely on the availability of human-labelled data that was originally not intended for evaluation purposes. One challenge of these existing domain-specific collections, is that they are released with no defined retrieval task. This makes systems often difficult to compare, because their training data is not the same, and\revision{, when the training data is a (web-crawled) sample of a larger dataset, its creation cannot be reproduced without considerable difficulty. }
A core example is the patent domain (see Section~\ref{sec:patents}), where human-assigned prior art -- \revision{previous patent applications or }previously granted patents -- are used as labels in evaluating patent-to-patent retrieval tasks~\citep{piroi2011clef}. Similarly, in the scientific domain, indirect relevance labels have been deduced from citations and co-citations (see Section~\ref{sec:bibliometrics}), for benchmarking query-by-document retrieval of scientific articles~\citep{cohan2020specter}. 

In 2018, the report of the first international workshop on professional search, concluded that:
\begin{quote}``The academic field of professional search is in need for benchmark data, as well as user observation studies and log data. The data should reflect the actual search problems and the topics should be temporarily grounded. Some of the useful, rich, collections are relatively small.  [...] For some purposes this is sufficient, but for quantitative system comparisons we need more topics, and a diversity of topics. As for now, it is still an open question what constitutes a sufficiently diverse dataset to evaluate our methods.''~\cite[p.160]{verberne2019first}\end{quote}
The situation has improved since then, especially with the release of the BEIR benchmark (see Section~\ref{sec:crossdomain}). But there is still a gap between benchmarking with static test collections for domain-specific search and evaluating search tasks in the context of work tasks (See Figure~\ref{fig:task_context} in Section~\ref{sec:model}).

\paragraph{Simulation studies} 
We follow the definition of~\citet{azzopardi_report_2011} that user simulations can be viewed as \textit{what-if} experiments: by simulating a range of user behaviour, we can evaluate how the search engine performs under the influence of specific behaviours. This is important because the perceived quality of IR systems highly depends on the behaviour of the users~\citep{maxwell2016simulating}. 

In the interaction with a search engine, a user will first decide to formulate an initial query. Since a poor query will not lead to relevant results, the content of the query matters for the effectiveness of the search. After the query has been issued, the user inspects the results, and the more persistent the user is, the more documents they will inspect and the more relevant information they will potentially gather. For example, if the most relevant document is ranked \revision{(i.e. placed) at position 8 in the result list, but the user only inspects the top~5 documents, the search will not be considered as successful. Consequently, in such situations, the Precision@10 score computed during an evaluation,} does not reflect the user's success experience. In addition to query processing and results ranking, search engines have interactive features, such as query suggestions. The effectiveness of the search directly depends on user choices in accepting query suggestions. 

User simulations are based on probabilistic models informed by real user observations~\citep{maxwell2016simulating}. For example, we can model the inspection behaviour of web search based on log data from a web search engine \citep{verberne_user_2015}: what is the probability that the user sees the snippet at position 5? And at position 20? Noise can be added to these probabilistic models to reflect unpredictable user behaviour~\citep{hofmann2011balancing}. It is good to note here that realistic models of user behaviour quite quickly become complex. For example, in the complex searcher model, \citet{maxwell2016simulating} define the choices that the user has in order to decide when to stop searching, and this alone involves extensive possibilities. A simpler approach is to assign each user action a time limit (e.g. 3 seconds for query formulation) and define user types by the time budget that they have for their search task~\citep{baskaya2012time}. 

While user models might not be fully representative of real users, these models are a way to reflect the range of user interactions with the system and systematically measure the effect of the user's behaviour on the search effectiveness.

\chapter{Retrieval models and methods}
\label{chap:retrieval_models_methods}


We start with a general introduction to two-stage retrieval, followed by methods for domain transfer and query processing. We then follow up with methods more specifically developed to address challenges in domain-specific search: models that can deal with long queries and long documents, the use of bibliometric data, and efficient (low-cost) models. The last two sections are devoted to domain-specific IR methods beyond document retrieval: conversational search and data augmentation.


\section{Two-stage retrieval and domain-specific encoders}
\label{sec:twostage}
Since the rise of BERT models for Natural Language Processing \citep{devlin2019bert}, the IR field has adopted transformer encoder models for ranking~\citep{dai2019deeper,nogueira2019passage,10.1145/3397271.3401325,lin2022pretrained}. The first approach to a BERT-based ranking was the monoBERT model~\citep{nogueira2019passage}. MonoBERT is a \textit{cross-encoder}: the query is concatenated with a candidate document and the relevance relation of the pair is learned in a supervised manner on labelled training data. At inference time, a new pair is concatenated, fed to the trained model, and a relevance prediction is made.

This approach, however, is computationally not feasible at inference time, as it requires to concatenate each document in the index to the query and, then, measure the relevance according to the trained model. This would entail millions of prediction computations by the transformer encoder while the user is waiting, with the result that the query latency is much higher than acceptable for a search engine. Therefore, an alternative and much more efficient approach to use BERT embeddings for semantic matching is the \textit{bi-encoder} architecture, first introduced by \citet{reimers-gurevych-2019-sentence} under the name Sentence\nobreakdash-BERT (SBERT). With bi-encoder models is is possible to embed all documents at index time and use the principle of \revision{Approximate Nearest Neighbour search} at inference time to find the most similar documents in a collection given a query. This type of retrieval models are called \textit{dense retrievers}.

The typical approach used in current search engines is a two-stage retrieval set-up. \revision{First, a highly efficient retrieval model is used to obtain the top-$k$ most relevant document from a large collection. This phase is termed as \textit{the first-stage retrieval}. 
In the second stage}, an effective re-ranker that was trained offline (e.g. monoBERT) is applied to re-rank these top-$k$ retrieved documents. Both lexical rankers such as BM25, and dense retrieval models such as DPR~\citep{karpukhin2020dense} and ANCE~\citep{xiong2020approximate} are commonly used for first-stage retrieval.

One of the benefits of BERT-based models is that they can be further pre\nobreakdash-trained on specific domains: starting from a regular BERT model, it can be made to be better suited for domain\nobreakdash-specific texts by feeding it a collection of domain\nobreakdash-specific texts with self\nobreakdash-supervised training. This process adapts the weights of the transformer encoder to the semantics of the added domain data. Many domain\nobreakdash-specific BERT models have been pre\nobreakdash-trained since 2019, most notably BioBERT~\citep{lee2020biobert}, PubMedBERT~\citep{gu2021domain},  LegalBERT~\citep{chalkidis2020legal}, PatentBERT~\citep{lee2019patentbert}, SciBERT~\citep{beltagy2019scibert}. Domain-specific non-English models are less common~\citep{carrino2021biomedical,brandsen2022bert,buonocore2023localizing}. 

We need to stress here that having a domain-specific BERT model is not synonymous to having a domain-specific ranker. To perform well on relevance ranking, BERT models have to be fine-tuned on pairs of queries and documents. This can be done either using a cross-encoder or a bi-encoder architecture. One of the challenges of cross-encoders, like MonoBERT, is that they need a large amount training data to outperform unsupervised rankers such as BM25 \citep{nogueira2020document}. This makes cross-encoders suboptimal for domains with limited training data available. As a result, ranking in specific domains is often done with a ranker that was not fine-tuned for the specific domain but instead on general-domain MS MARCO data.

\section{Methods for domain transfer}
The general goal of domain adaptation for ranking models is to enable effective training on a smaller domain-specific dataset after a model was trained on a much larger dataset from a resource-rich domain. The adaptation can be seen as a machine learning problem where the model weights are not trained from scratch but adapted based on the smaller dataset~\citep{geng2009ranking}. 

In the common transfer learning paradigm with transformers, the model is pre-trained in a self-supervised way on a large amount of unlabelled data and then fine-tuned for a specific task on a labelled data set~\citep{devlin2019bert}. Starting from the same pre-trained BERT model, supervised fine-tuning for multiple tasks is possible: for text classification, labels are added on the document level; for named entity recognition, labels are added on the token level~\citep{devlin2019bert}. If the target domain is very different from the one on which the base model was trained, inserting domain vocabulary may help improve the quality of the model~\citep{yeung2019effects,gu2021domain}.

In the case of relevance ranking as a supervised task, a BERT model has to be fine-tuned on pairs of queries and documents. If there is sufficient data available, a pre\nobreakdash-trained BERT model can be fine\nobreakdash-tuned in a cross\nobreakdash-encoder setting on pairs of queries and documents, e.g. the Chinese BERT-large was fine-tuned on the LeCaRD dataset (see Section~\ref{sec:legal}) by \citet{zhu2022bert}. Since for many domains there are not sufficient labelled query--document pairs available to effectively train a domain-specific ranker (see Section~\ref{sec:twostage}), the common fine-tuning setup for transformer-based rankers consists of two steps: A pre-trained model is first fine-tuned for the \textit{relevance ranking} task on a general-domain dataset (typically MS MARCO), and then further fine-tuned on smaller labelled data for the specific topic domain~\citep{weber2024comparative}. In the first fine-tuning stage the model learns the relevance relation between queries and documents, while in the second fine-tuning stage it further specifies the learned relevance relation to the specific domain at hand. When previously trained models are used on a new dataset without fine-tuning (e.g. a model trained on MS Marco applied to a medical domain ranking problem), we refer to this as \textit{zero-shot use}. For both bi- and cross-encoders pre-trained models are available\footnote{\url{https://sbert.net/docs/pretrained_models.html}}. 

Another approach to domain adaptation is the use of data augmentation~\citep{bonifacio2022inpars}. This is discussed in more detail in Section~\ref{sec:augmentation}.

\section{Query processing}\label{sec:queryproc}
Queries in domain-specific tasks are typically more specific and often more complex than in the general (web) domain (see Section~\ref{sec:complex_needs}), owing to the information needs being user- and context-specific~\citep{mason_legal_2006}. In this section, we focus on the challenges of Boolean queries in particular (writing and processing), and describe approaches to deal with query ambiguity and domain specific query interfaces.

\paragraph{Boolean query writing} In some knowledge-intensive tasks, Boolean queries are common. Examples are patent prior art search, systematic reviewing, and news monitoring. Creating Boolean queries is a challenging activity: it is common that the Boolean query for a particular systematic review task consists of dozens of components (clauses). Even experienced professional information specialists may have challenges in creating good Boolean queries~\citep{verberne2016boolean}. It is, therefore, useful for experts to have some assistance, when formulating new or adapting existing queries. This motivates research on automatic Boolean query generation or suggestion, especially in the context of systematic reviews \citep{scells2018generating,scells2019automatic,scells2021comparison,wang2023can}. \citet{scells2018generating} define possible transformations of Boolean queries -- of the the AND and OR clauses inside a query -- that lead to better retrieval results than the original human queries. In a follow-up study, \citet{scells2018generating} proposed automatic query refinement strategies, aimed to assist the professional searcher formulate better queries. After the rise of Large Language Models (LLMs), the same group of researchers experimented with ChatGPT for Boolean query generation~\citep{wang2023can}. They find that, when instructed well, the LLM has potential for query formulation in the context of systematic reviews, especially in settings where time is limited and higher precision is prioritized over high recall. One of the challenges identified is that ChatGPT sometimes generates non-existing vocabulary (MeSH) terms. Solving this requires more guided uses of generative models.

\paragraph{Boolean query processing} Boolean queries are not only challenging for users to formulate, they can also be complex for search engines to process. Boolean queries were originally used in a strictly Boolean retrieval context: any document in the collection would be matched to the Boolean pattern and the set of matching documents was returned. In long and complex queries, this has the effect that one additional OR statement may heavily increase the set of retrieved documents, while one additional AND statement may reduce the set of returned documents to zero. Current-day search engines that allow Boolean queries for complex search tasks (e.g. systematic reviewing) combine the specificity of Boolean operators in the query with relevance ranking. This is not a trivial undertaking, since the AND and OR Boolean operators do not naturally constitute a relevance ranking. As a solution, \citet{scells2020you} propose to use rank fusion to the clauses of a Boolean query, where the intuition is that documents that are returned by many clauses are more likely to be relevant.
\paragraph{Query ambiguity} A challenge in many IR contexts is that user queries tend to be ambiguous with respect to the underlying information need. In web search, queries are typically short \revision{(a few terms)} and can have multiple different interpretations depending on who asks it and when, e.g. the query ``einstein leiden'' can refer to either a visit of Albert Einstein to Leiden or to the cafe called Einstein in Leiden. Domain-specific queries are generally longer than open-domain web search queries, but may also suffer from ambiguity. For example, the query ``search behaviour'' has a different meaning in the context of the information sciences than it has in the context of predator biology. For that reason, methods for query disambiguation have been proposed in domain-specific contexts. Approaches to query disambiguation can be based on domain information~\citep{bonab2019incorporating} or query expansion (query augmentation with tags~\citep{broder2008domain}). In the patent domain, the information need is represented by a seed patent document (see Section~\ref{sec:qbd}) and the patent specialist needs to find good query terms to formulate an effective Boolean query. Finding good query terms requires disambiguation of terms based on semantic similarity to prevent query drift~\citep{mahdabi2013leveraging}. 
Another common solution \revision{that avoids query disambiguation} is \textit{diversification}, which is showing results for different meanings of the query on the search engine result page. Web search engines use this approach actively, showing a large variety of results for any given query. In the legal domain, diversification can be improved with domain-specific criteria such as \revision{Topical taxonomies and Time similarities as described by }\citet{koniaris2016multi}.
\paragraph{Query interfaces} Query processing is directly related to query user interfaces. One common query creation aspect in domain-specific search is the use of facets: in a faceted search engine, the user can select metadata filters to, e.g., govern the year of publication or the document type \citep{tunkelang2009faceted}. Pubmed is a typical example of a faceted search system~\citep{tang2007browsing}. Domain-specific user groups might require more specified query interfaces and facets, such as a time scale for historical search tasks~\citep{verbernea2017rembench} and a map for archaeological search~\citep{brandsen2019user}.
\section{Dealing with long queries and long documents}\label{sec:dealingwithlong}
As introduced in Section~\ref{sec:long}, long documents pose challenges to ranking models. In this context, we define `long' as too long to fit in the input of a regular BERT model, which is limited at 512 tokens. Roughly speaking there are four different approaches to deal with long documents. 

\paragraph{Truncating} The most straightforward and most used approach is truncating the input \revision{to use only the first 512 tokens}~\citep{lin2022pretrained,abolghasemi2022improving}. For many document types, this does not reduce the effectiveness of the model too much because the most important information is in the beginning of the document. This is particularly the case for genres where documents starts with a title and abstract (scientific documents, patents, news). \revision{However, truncating is not a good strategy when handling} more specific information needs where information from later sections is required, e.g. queries related to \revision{detailed invention descriptions in patents or patent claims}, which are generally towards the end of the documents. Truncating is problematic also for documents in which the information is evenly distributed, such as Wikipedia articles. For case law, \citet{ma2022incorporating} looked into finding the best cut-off truncation location.

\paragraph{Use of lexical models} Compared to BERT-based rankers, BM25 can deal with any input length, and \revision{BM25 variants that account for large document length divergence readily exist}~\citep{kamphuis2020bm25}. It was shown that for domain-specific collections, which are often relatively small, it is difficult to beat the BM25 baseline with BERT-based models~\citep{askari2021combining}.

\paragraph{Splitting into shorter segments} When truncating collection documents would lead to information loss, the alternative is to split them into short segments of less than 100 tokens and select the most relevant ones from each document given an input query. The most relevant segments are then concatenated into a pseudo-document that is used for relevance ranking \citep{li2021keybld,li2023power}. This approach can also be combined with rankers that apply so-called late interaction between a query and a document. Late interaction as proposed by~\citet{khattab2020colbert} in their ColBERT model embeds the query and document separately (like a bi-encoder). At index time, the model computes the dot product between each query term and each document term, identifies the most similar document term for each query term, and sums over these similarities. \citet{li2022bert} apply this principle to long documents, using pre-stored term embeddings to compute passage scores, which are then aggregated into one document score.

\paragraph{Automatic summarization} A viable alternative to splitting documents into shorter segments is summarizing the document into a shorter pseudo-document~\citep{bhopale2024transformer,askari2021combining,bijl2024efficient}. This has two advantages: First, as explained above, the shorter document can be more easily processed by transformer-based models that are limited in their input length by memory burden. Second, it potentially reduces the amount of noise in the document, which makes the shorter document more informative and relevant to the retrieval task. This approach requires the availability of high-quality summarization models that can take long inputs -- i.e. long-context LLMs, or LongFormer-based sequence-to-sequence models~\citep{beltagy2020longformer} -- and have been trained for the domain at hand.

\begin{figure}[t]
    \centering
    \includegraphics[width=1.0\linewidth]{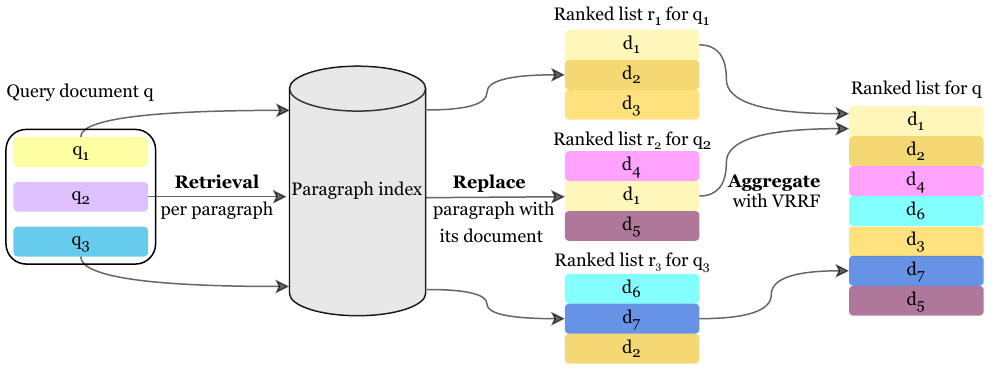}
        \caption{Paragraph-level document retrieval with aggregation with additive integer scores to an overall ranked list for the query document $q$ (from \citet{althammer2022parm})}
        \label{fig:paragraphlevelretrieval}
\end{figure}

\paragraph{Paragraph or sentence-level aggregation} In query-by-document (QBD) retrieval tasks, such as patent prior art retrieval or case law retrieval, special attention is required in processing the query document. Since the information need might be spread over the complete input document, truncating is often not a good idea. Lexical models are sometimes used in combination with keyword extraction: instead of feeding the whole document to the ranking model, extract, first, the most salient terms from the query document, and use those as a pseudo-query \citep{locke2017automatic,askari2021combining}. Another solution when using transformer models to process long query documents is paragraph- or sentence\nobreakdash-level retrieval and aggregation \citep{yang2019simple,li2020parade,zhang2021comparing,althammer2022parm,askari2024retrieval}. In this approach, the query document is split into smaller segments, and for each segment the top-$n$ relevant segments are retrieved from the document collection. In order to create a ranked list of documents, the relevance of paragraphs is interpolated to document relevance through aggregation. This is illustrated in Figure~\ref{fig:paragraphlevelretrieval}. 

\section{Vertical search} \label{sec:vertical}

\revision{In the context of web search, a \textit{vertical} is one particular segment (or index) of the web from which results are retrieved. Web search engines aggregate search results from multiple different index partitions in a single interface. These verticals are generally based on modality (images, videos, maps, news), topicality, publication source, date, or domain \citep{zhou2011evaluating,zhou2013vertical}.}

\revision{\citet{Lewandowski2023} defines vertical search as the specialization of information retrieval systems toward particular collections, where indexing, ranking, and interface design are tailored to the needs of a given community, such as medical or legal experts \citep{shettar2008vertical}. This makes vertical search particularly relevant to domain-specific IR. The idea is that by narrowing the scope of the search space, vertical search systems can achieve higher precision, especially in domains where terminology, metadata, and trustworthiness are critical.}

\revision{In the healthcare domain, vertical search can be an approach to access multiple types of information, including archived question--answer pairs, biomedical vocabularies, and structured records, through one interface \citep{nie2014wenzher}.  More recent work explores domain-specific pretraining for biomedical vertical search, addressing challenges of information overload and specialized terminology in biomedical collections \citep{wang2021domain}. }

\revision{As \citet{Lewandowski2023} notes, vertical search engines are increasingly integrated into broader web search environments like Google, but in high-stakes contexts -- systematic reviewing, case law -- vertical search provides more control over the search process, selecting trustworthy index partitions matching the information need at hand.}

\section{Using bibliometric data (citations)} \label{sec:bibliometrics}
In some domains, citations to related documents play an important role for retrieving relevant information~\citep{mayr2014bibliometric}. The most striking example is the academic domain, with extensive literature references between papers~\citep{cabanac2020bibliometric}. Already in 1963, \citet{salton1963associative} proposed the use of citation information to retrieve associated documents. Before the rise of web search engines, citations were used by librarians and scholars to identify related, relevant documents. In the early work on academic information retrieval, citations were therefore considered even more important than keyword search: finding relevant papers through `snowballing' from one paper to the cited papers was a common approach for literature search. 
\citet{ellis_modeling_1993} calls this \textit{chaining} \revision{and defines it as} ``following chains of citations or other forms of referential connection between material'' \citep[p.482]{ellis_modeling_1993}. Still, snowballing is used for scientific reviewing~\citep{choong2014automatic}. Citation references can be used not only by users for chaining, but also by search systems for clustering. Document clusters can improve the user experience in academic search and browsing~\citep{cobo2011science,van2011methodological,bascur2023academic}. The context of citations in scientific documents can be used as expansions of indexed documents, e.g. by appending citation contexts from citing documents to cited documents \citep{dabrowska2015exploiting}.

The use of bibliometric information\footnote{Three closely related terms referring to the quantified use of citation references are scientometrics, bibliometrics, and informetrics. The terms are used in different context, but are quite similar in their concepts and methods~\citep{yang2020scientometrics}. We use the term bibliometrics here.} in search engines for academic information retrieval has been addressed in the Bibliometric-enhanced IR (BIR) workshop series \citep{mayr2014bibliometric,cabanac2020bibliometric}. Over the years -- the workshop has had 14 editions up to 2024 -- central topics in the workshop have been academic search and recommendation systems, scientific literature mining, and user needs in academic search. In 2025, the BIR workshop has been re-cast into SCOLIA, \textit{Scholarly Information Access}. SCOLIA has a broader scope than BIR, incorporating the growing importance of generative AI and NLP for the Scientometrics and \revision{academic/scholarly} IR communities.  

The importance of citations extends from the academic to the biomedical domain, but also to the legal \citep{wiggers2023bibliometric} and patent domains~\citep{shalaby2019patent}. In all these domains, citations and co-citations can indicate relatedness between documents and can be used as pseudo-labels to train relevance models~\citep{cohan2020specter}, especially for query-by-document retrieval~\citep{abolghasemi2022interpolation}. The idea is that if document A is cited from document B, then A is relevant to B, and if A and B are both cited from C, then A and B are relevant to each other. In patent retrieval benchmarks such as CLEF\nobreakdash-IP, such citations are considered ground truth relevance labels for patent prior art retrieval~\citep{piroi2011clef}. One limitation of this approach is that citation information cannot be used as labels and features at the same time. For example, a retrieval method that uses citation-based document clustering~\citep{bascur2023academic} cannot be evaluated against relevance labels obtained from citations or co-citations.

\section{More efficient methods (lower computational burden)}

For search engines, inference speed is of great importance. The waiting time for a user to receive the results for their query (called \textit{latency}) affects user behaviour~\citep{arapakis2014impact}. This is why in the development of ranking models, optimization of ranking quality is not an only goal: inference speed cannot be sacrificed for the sake of a better ranking. Due to privacy restrictions in some domain-specific search applications, \revision{like patient data which must be searched within a hospital's internal infrastructure, or like the prefiling patent search that a company does and which must not, potentially, reveal to competitors its current technological developments}, domain-specific search applications sometimes have requirements to run on in-house hardware. As such hardware can be less performant than what would be available on the cloud, efficiency is particularly important for these domain-specific search applications. In this section we discuss approaches to make ranking efficient.

\begin{figure}[t]
    \centering
    \includegraphics[width=1.0\linewidth]{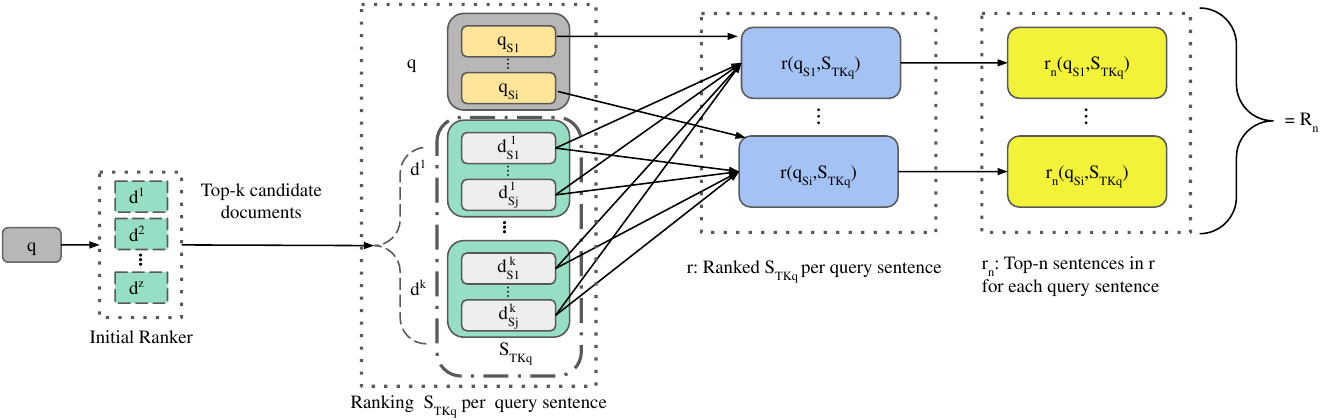}
        \caption{Processing long queries and documents efficiently by finding the most similar sentences between a query document and a candidate document (from \citet{askari2024retrieval})}
        \label{fig:rprs}
\end{figure}

\paragraph{Efficient first-stage ranking} As introduced in Section~\ref{sec:twostage}, two-stage retrieval is the most viable strategy to balance speed and quality: in the first stage, a set of documents is retrieved from the (potentially huge) index using a very efficient indexing and search strategy. Traditionally, this is implemented as an inverted index combined with a lexical retrieval model. In the second stage, a more powerful ranker (e.g., a cross-encoder) will re\nobreakdash-rank the much smaller set of documents obtained in the first\nobreakdash-stage. One potential problem of lexical matching in the first-stage retrieval is that relevant documents that do not have a lexical overlap with the query, are not retrieved from the index, and cannot be ranked in a high position in the second stage. For that reason, for first-stage retrieval, transformer-based bi-encoders were proposed as a good alternative to BM25~\citep{reimers-gurevych-2019-sentence,karpukhin2020dense}. In a bi-encoder architecture documents are embedded at index time, and, at query time, the model efficiently finds the most similar document in the dense, pre-trained vector space using Approximate Nearest Neighbours (ANN) search~\citep{xiong2020approximate}. ColBERT \citep{khattab2020colbert} is a hybrid implementation, combining the effectiveness of a cross-encoder architecture (interaction between query terms and document terms) and a bi-encoder architecture (ANN). The model is highly efficient and can be used for first-stage retrieval as well as re-ranking. In domain-specific IR, the longer documents can negatively impact the ranking efficiency even more (also see Section~\ref{sec:long}). Similarly to the late interaction principle in ColBERT, extremely long queries and documents can be processed efficiently using the most similar sentences between a query document and a candidate document (see Figure~\ref{fig:rprs}). 

\paragraph{Document expansion} Another way to benefit from the efficiency of lexical rankers while including semantic matching is to expand the document during index time with semantically related terms that are potentially used in a query for which the document is relevant. \revision{This is a form of data augmentation discussed in Section~\ref{sec:augmentation}}. 

\paragraph{Reducing the model size} Transformer-based models can be made more efficient by reducing their size (number of parameters). BERT and RankT5~\citep{zhuang2023rankt5} have several model sizes where the smallest models need less computational memory and \revision{have shorter inference times}. A model size can be reduced by \textit{quantization} techniques: reducing the \revision{parameter values' precision} from 16 or 32 bits floating point to 8 or even 4 bits. This reduces the memory footprint and increases the model's processing speed~\citep{bijl2024efficient,yang2022compact}.

\section{Domain-specific conversational search}


Conversational information seeking~\citep{zamani2023conversational} is a form of IR in which the user does not enter one keyword-based query to retrieve a single list of documents, but goes into a sequence of interactions with the search engine to collect relevant information. Research in conversational search addresses four main tasks: query reformulation, search
clarification, answer retrieval, and response generation~\citep{mo2024survey}. In this section we focus on retrieval methods.

The queries in conversational search are typically formulated as natural language questions instead of the short queries in conventional web search engines. This aspect makes conversational search related to question answering (QA). In conversational search, effectiveness can be measured not only for individual requests but also for the session as a whole. In that regard conversational search is related to session-based search and exploratory search. Traditionally, there are three different approaches to conversational search: 
\begin{itemize}[itemsep=0pt, parsep=0pt, partopsep=0pt]
    \item \textbf{Retrieval-based methods}: select the best possible response from a collection of responses with the use of indexing and retrieval techniques), 
    \item \textbf{Generation-based methods}: generate a response in natural language with the use of generative language models, 
    \item \textbf{Hybrid (Retrieval Augmented Generation) methods}: first retrieve potentially relevant information from an index, then generate the response based on the retrieved information. 
\end{itemize}
    
\paragraph{Retrieval-based methods} The main advantage of these approaches is that search engine responses come directly from the index, and are, thereby, transparent and reliable. This is advantageous for both laypeople and expert use cases.

In a professional legal context, \citet{liu2021conversational} show that 
users can achieve better retrieval effectiveness with a conversational legal case retrieval system than with traditional search. For the use of laypeople, models for passage retrieval in conversational search can be trained or fine-tuned on question-answer pairs or on informational conversations between humans, e.g. legal or health forum posts where experts answer questions of laypersons. This approach was applied to the legal context by \citet{askari2023closer,askari2024answer}. In the \revision{human\nobreakdash-generated, original data, questions are asked by lay people, while} the answers are written by lawyers (experts), and there is a conversation between a lay person asking a question, and one or more lawyers providing information. In critical contexts like these, if we want to partially replace the experts by automation, it is a safe option to restrict the output to answers already written by experts before. 

In conversational search, the user expects the search engine to remember the dialogue context and be able to formulate responses that take \revision{into account the previous questions and queries. It is non-trivial to achieve this in a retrieval setting, since the transformer encoder requires to be fed, at each conversation step, not only the query but also a selection of the previous dialogues for which it} has to learn the relevance of indexed response passages. 

The disadvantage of the retrieval-based approach is that there may be no question that is highly similar to a newly incoming question, or that the relevant answer has to be compiled from multiple existing answer passages. 
\paragraph{Generation-based methods} With the capability of LLMs to generate fluent text, generative conversational approaches have become a viable choice. Even more, in the current information-seeking landscape, the use of ChatGPT for information finding is very common among the general public ~\citep{dash2023evaluation,zhao2023can,mcgee2023using,xu2023chatgpt}. For example, it was shown that, in some cases, patients preferred the ChatGPT output over that of a medical doctor~\citep{ayers2023comparing}. ChatGPT and other generative LLMs that have been fine-tuned for conversation are capable of answering follow-up questions by using the full context of the dialogue as prompt for generating the next response\revision{~\citep{roumeliotis2023chatgpt}}. A purely generative approach however is not viable in expert domains, because the hallucination rate of generative LLMs increases steadily when the topic becomes more specific. It was shown that ChatGPT’s reliability for answering questions is the lowest in the law and science domains~\citep{shen2023chatgpt}. Another study in the medical domain showed that less than 20\% of the responses of GPT-3.5 and GPT-4 for information finding in healthcare agreed with expert answers, and that the responses contained hallucinated references~\citep{dash2023evaluation}. Although hallucination can be reduced with smart prompting strategies~\citep{dhuliawala2023chainofverification}, hallucination can never be fully prevented because of the nature of probabilistic text generation: texts are generated word by word. Also, without references to sources, the user cannot verify the output of the conversational LLM, which is problematic in domain-specific contexts where details matter.

\paragraph{\revision{Hybrid, Retrieval-Augmented Generation (RAG) methods}}
To deal with the issues mentioned above, the most common approach to conversational search nowadays is a hybrid approach, and in particular Retrieval-Augmented Generation (RAG) approaches~\citep{NEURIPS2020_6b493230}. 

In RAG, the retrieval component provides sources for a generative LLM to base its answer on. RAG brings several advantages to generative models: it reduces hallucination~\citep{shuster2021retrieval}, it makes the source of the generated text transparent to the user, and it enables the generation of responses for specific, pre-defined collections, like private (company-owned) collections. RAG combines the strength of retrieval (being source-transparent and reliable) with the strength of generative LLMs (being interactive and fluent). In addition, through the specification of the potentially relevant documents in the context of the LLM, RAG enables domain adaptation~\citep{siriwardhana2023improving}. A good number of studies since 2023 have addressed RAG for specific domains: healthcare~\citep{wang2023potential}, financial~\citep{iaroshev2024evaluating}, and legal~\citep{redelaar2024attributed}. Note that not all these studies are conversational search approaches in the strict sense, because they do not always study and evaluate conversations with multiple turns (follow-up questions). Studies addressing RAG for multi-turn conversations in specific domains are much sparser and there is much to gain through research in that direction. The relation between domain-specific RAG and IR is discussed further in Section~\ref{sec:RAG}.

\section{Data augmentation} \label{sec:augmentation}
There are two main purposes for data augmentation in IR. The first is to augment documents in the index for improving the relevance mapping between queries and documents. The second is to create more training data in contexts where labelled data is lacking. In this section, we discuss both.

\paragraph{Bridging the vocabulary gap} The vocabulary gap is a long-standing issue in IR: the query might contain different words than the relevant documents in the collection~\citep{muller2008closing}. This problem is predominant in lexical retrieval models. For example, when searching with the query ``bike path'', a document about ``bicycle paths'' would be relevant, even though none of the terms overlap. Dense retrieval models alleviate this problem, but are less efficient than lexical models.

One option to bridge the vocabulary gap without losing efficiency at query time is to expand documents with potential query terms during index time. A common method for this is doc2query and its successor docT5query~\citep{nogueira2019doc2query}. The idea of this approach is that a sequence-to-sequence model is trained on pairs of queries and relevant documents that, given a document, produces potential queries for which the document would be relevant. These potential query terms are then added to the document in the index. DocT5query uses a transformer\nobreakdash-based model (T5) that is capable of generating semantically similar query terms.

\paragraph{Creating more training data} In domain-specific contexts it is sometimes necessary to create additional query--document pairs \revision{for model training. The main reason is that, due to the high costs of labelling by domain experts, there is a lack of labelled data for training models~\citep{althammer2022tripjudge}. For training data-hungry ranking models like BERT, expert labelled data is often sparse (see Section~\ref{sec:twostage}).} \citet{lin2022pretrained} show that a BERT cross-encoder can only improve over BM25 when there are more than 5000 relevant query–passage training instances~\citep[Fig.~8]{lin2022pretrained}. In specific domains, the shortness of training data is aggravated because the information needs tend to be user-specific and have a long-tail distribution (see Section~\ref{sec:eval}).

Thus, in domains with limited training data, it can be valuable to use automated methods to expand it. In many scenarios, there will be an abundance of unlabelled data: often a document collection and in some cases query logs or questions. If a document collection is available but no queries, a generative model can be used to generate synthetic queries from documents~\citep{ma2021zero}. A commonly adopted method is InPars~\citep{bonifacio2022inpars,jeronymo2023inpars}. The assumption is that with a properly trained sequence-to-sequence model on document\nobreakdash-query pairs, or a well\nobreakdash-instructed generative LLM, the input document will be relevant for the generated queries. Generative Pseudo Labelling (GPL) combines a T5-based query generator with pseudo-labelling with a strong retrieval model to create not only relevant query-document pairs, but also non\nobreakdash-relevant documents that are similar to the relevant documents \citep{wang2022gpl}. If a collection of queries or questions is available, an option is to generate synthetic documents (answer passages, typically) for queries~\citep{askari2023generating,askari2023expand}. Starting from real user queries instead of documents could ensure more diversity in the training collection, better reflecting the long tail of user needs. On the other hand, synthetically generated documents may not be completely representative of real documents in the collection.

Since 2023, there has been a discussion in the IR community about using LLMs for relevance assessment~\citep{faggioli2023perspectives}. Consensus is that LLMs can be used for augmenting human-created data, and that the best is to combine training data from multiple sources~\citep{faggioli2024determines}. In the context of domain-specific IR, this use of LLMs appears to show potential since it allows training of ranking models on more data while retaining the traditional search engine connection to actual information sources -- an important aspect for professional users who require control of the search process and trustworthiness of sources. This will be further discussed in Section~\ref{sec:genLLMs}.

\chapter{Domain-specific IR in the age of LLMs}
\label{chap:LLMs}

Since 2022, LLM-based chatbots have become ubiquitous. Models such as OpenAI's GPT-4, Microsoft's Copilot, and DeepSeek have been optimized for conversations, enabling natural language interactions to solve problems. These generative models are used in countless domain-specific contexts: as writing assistant~\citep{basic2023better}, programming assistant~\citep{surameery2023use,tian2023chatgpt}, and for information finding~\citep{dash2023evaluation,zhao2023can,mcgee2023using,xu2023chatgpt}.

\section{The potential of IR for generative LLMs}\label{sec:RAG}
\paragraph{The problem of generative LLMs for information finding} 
Generative AI has a risk of generating false information; a problem commonly referred to as \textit{hallucination}. Hallucination has been studied and quantified extensively since 2023, showing that a substantial proportion of the output generated by LLMs across a large range of tasks is hallucinated~\citep{huang2025survey,zhang2023siren,ji2023survey}. The hallucination problem is the most severe on topics that have relatively low-frequency in the pre-training data. \citet{shen2023chatgpt} show that for answering questions, ChatGPT’s reliability is the lowest in the law and science domains. Sycophancy -- the tendency of conversational LLMs to admit mistakes and agree with the user -- is causing even more misinformation with users who do not have the correct domain knowledge~\citep{liu2023trustworthy,park2023ai}.

When used for information finding, hallucination is a serious problem, especially because users are not always aware of the probabilistic design of the model and trust its output. With a study among 607 active users of ChatGPT in the US in 2023~\citep{choudhury2023investigating}, it was found that 36\% of them use ChatGPT for information gathering. A smaller 2025 study \citep{yazan2025personality} showed that daily users of both ChatGPT and Google have higher trust in ChatGPT than in Google, perceiving it as more human-like, and express willingness to trade factual accuracy for enhanced personalization and conversational flow. Unfortunately, it is common for users of LLMs to trust the model output~\citep{xu2023chatgpt,choudhury2023investigating}. This can especially be harmful in contexts where information from LLMs is used to make decisions, for example in healthcare~\citep{sallam2023utility,dash2023evaluation} or psychotherapy~\citep{sedlakova2023conversational}. 

In domain-specific tasks, to verify the quality of the information provided by a generative model, it is important that (1) the user is sufficiently knowledgeable to judge the correctness of the generated information and (2) that the answer provided by the chatbot is grounded in sources. This latter goal is addressed by Retrieval-Augmented Generation (RAG).

\paragraph{Retrieval-augmented generation (RAG)} 
RAG was proposed in 2020 \citep{NEURIPS2020_6b493230} to deal with so-called knowledge-intensive language tasks (KILT)~\citep{petroni2021kilt}. In RAG, a model first retrieves potentially relevant documents or passages from an index, then feeds it to the LLM together with the user query to generate a response that is based on the retrieved information. RAG has been shown to reduce hallucination~\citep{shuster2021retrieval}. \revision{There are further advantages of RAG over merely LLM use}: 

\begin{enumerate}[label=(\roman*)] \item RAG can attribute the generated content to a source (by referencing the source document)~\citep{qi2024model}, which allows the user to verify the information; 
\item The information can come from restricted sources like company-internal collections or even personal collections~\citep{ryan2025enronqa}. This allows for tasks such as enterprise question answering~\citep{yu2025ekrag,mukherjee2025documents}; 
\item The retrieval index can be updated without retraining the LLM. This enables question answering about recent topics, but also offers opportunities for domain adaptation \citep{xu2023retrieval} or even personalisation~\citep{salemi2024lamp,yazan2025improving}. 
\end{enumerate}

\begin{figure}[t]
    \centering
    \includegraphics[width=0.8\linewidth]{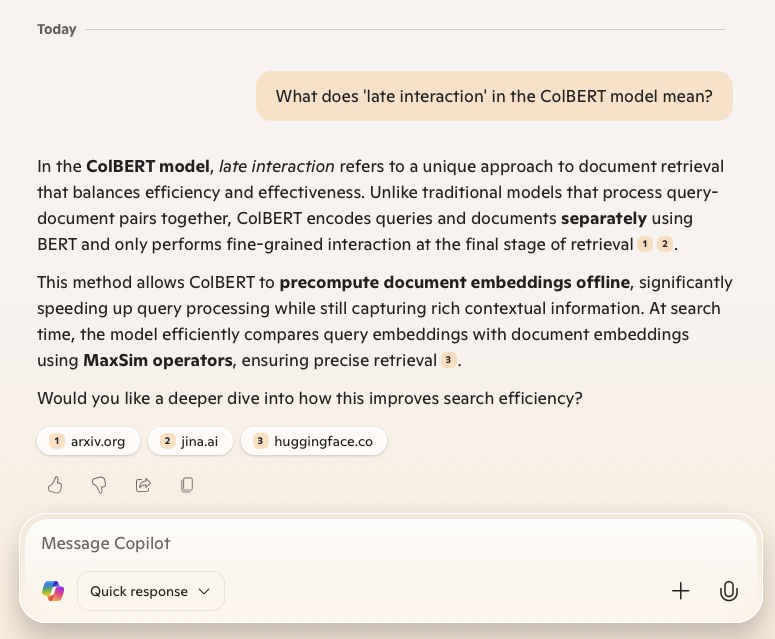}
    \caption{Example of Retrieval-Augmented Generation with answer attribution for academic Question Answering in Microsoft's Copilot (June 2025)}
    \label{fig:copilot}
\end{figure}

These advantages of RAG being more source-transparent and controllable than generative-only chatbots has led to a large wave of RAG approaches in domain-specific settings: in legal~\citep{wiratunga2024cbr,pipitone2024legalbench,hou2024clerc,louis2024interpretable,redelaar2024attributed}, biomedical~\citep{soman2024biomedical,matsumoto2024kragen}, and academic literature~\citep{aytar2024retrieval,bevara2025prospects,asai2024openscholar}. The common argument for using retrieval before generation in domains is that grounding of generated information in established knowledge sources is important to (1) prevent hallucination in critical contexts and (2) allow for verification of the information. At the time of writing, both OpenAI's ChatGPT and Microsoft's Copilot have built-in RAG functionality, providing URL references to relevant statements in the generated text. An example is shown in Figure~\ref{fig:copilot}.

\section{The potential of generative LLMs for domain-specific IR}\label{sec:genLLMs}

We have discussed the uses of generative LLMs for data augmentation in Section~\ref{sec:augmentation}. Generative models (typically LLMs) can be used to create or extend test collections for domain-specific IR tasks by either generating queries~\citep{jeronymo2023inpars}, generating documents~\citep{askari2023generating}, or generating relevance labels~\citep{faggioli2023perspectives}. In that line of work, the role of generative models is limited to the data creation phase; the actual retrieval and ranking is achieved by traditional rankers, term-based or embedding-based.

\begin{figure}[t]
    \centering
\includegraphics[width=0.8\linewidth]{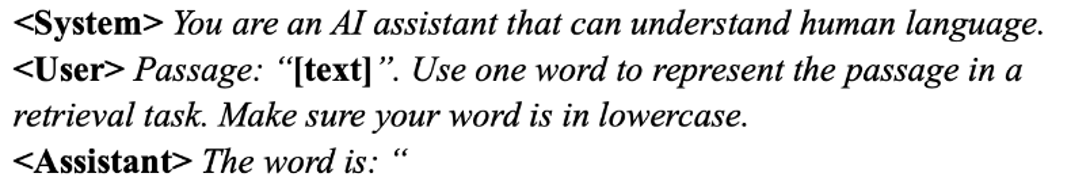}
    \caption{Prompt used by PromptReps~\citep{zhuang2024promptreps} to generate a single token with an embedding representative of the whole text (query or passage)} 
    \label{fig:promptreps}
\end{figure}

In this section, we briefly discuss three forms of generative IR, i.e. retrieval and ranking approaches with generative models as central component, and their applications to domain-specific contexts:

\paragraph{Using generative LLMs to generate embeddings} Generative LLMs are built on the transformer architecture~\citep{vaswani2017attention}. In a transformer, each token is represented by an embedding vector. It is common to use transformer \textit{encoder} models like BERT for embedding documents for the purpose of retrieval, but \textit{decoder} models like GPT or Llama also offer the potential of creating embeddings. This idea was investigated by~\citet{zhuang2024promptreps}, proposing a method called PromptReps. In PromptReps, an LLM is prompted to represent a text (query or passage) into a single token that represents the complete text (see Figure~\ref{fig:promptreps}). The embedding of the generated token is then indexed for Approximate Nearest Neighbour search, using a bi-encoder. In addition, PromptReps uses the top-$k$ generated tokens after the prompt to store lexical items in an inverted index. This leads to a hybrid ranking model, combining lexical and embedding-based matching. Evaluation on 13 test collections in multiple domains shows that the effectiveness of PromptReps depends on the test collection~\citep{zhuang2024promptreps}, but compared to encoder models that are trained with contrastive learning, PromptReps does not require any (contrastive) pre-training costs. This makes it a promising approach for domain-specific settings where labelled training data is sparse.

\paragraph{Prompt-based re-ranking} When instructed properly, LLMs can do relevance ranking of documents for a query. Note that this approach is not intended for retrieval or full ranking (on the complete document collection) but used in the re-ranking stage. Given a query pre-retrieved set of documents, the idea is to prompt the LLM for relevance, or directly for ranking. Four different prompt types are distinguished, adopted from learning-to-rank settings: \begin{itemize}[noitemsep]
\item \textbf{Pointwise}: getting a relevance estimate of a single query--document pair. A pointwise prompt is, given ``Passage: \texttt{\{text\}}. Query: \texttt{\{text\}}: ``Does the passage answer the query?''~\citep{qin2024large} or ``Is this passage relevant to the query?''~\citep{ma2023zero}. The LLM is answering the question with one of the two tokens `Yes' or `No'. The logits or probability that the generative model assigned to the `Yes' token can be used as a proxy for how certain the model is that the document is relevant; this is then used to rank the documents.
\item \textbf{Pairwise}~\citep{qin2024large}: getting a preference ranking between two documents for the same query. A pairwise prompt is ``Given a query \texttt{\{query\}}, which of the following two passages is more
relevant to the query?'' The LLM selects the more relevant documents from all pairs, and a full ranking is deduced from the pairwise preferences of the model, e.g. using Heapsort ~\citep{qin2024large}.
\item \textbf{Listwise}~\citep{ma2023zero,sun2023chatgpt}: getting a ranking of a list of retrieved documents. An example of a listwise prompt is shown in Figure~\ref{fig:listwiseprompt}. 
\begin{figure}[t]
    \centering
    \includegraphics[width=0.7\linewidth]{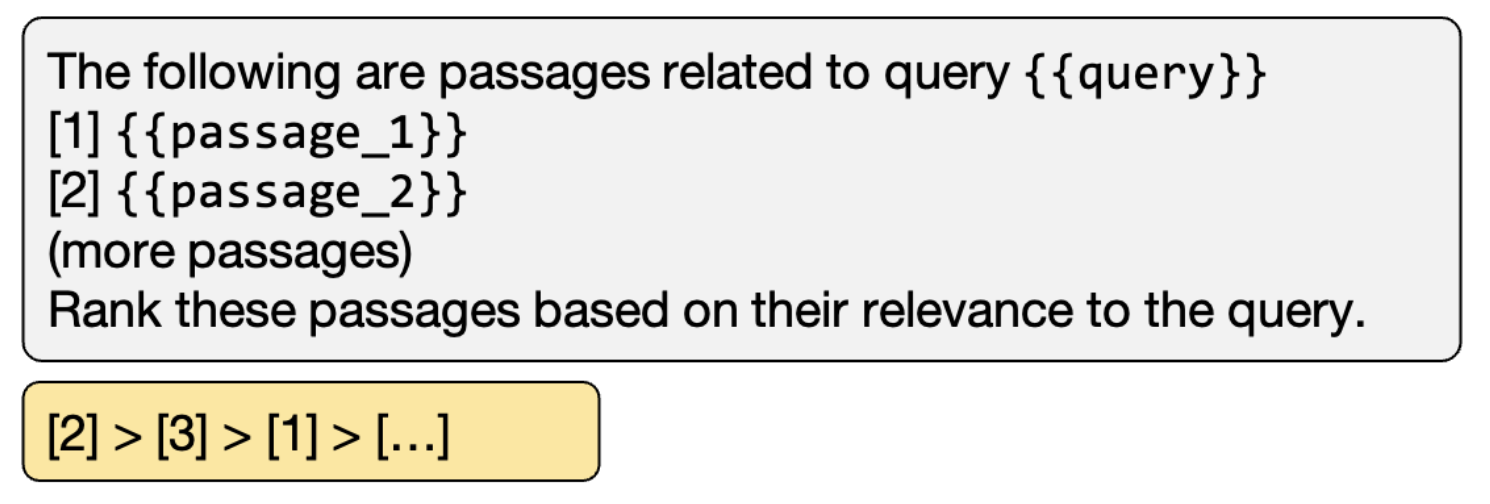}
    \caption{Example of a listwise ranking prompt~\citep{sun2023chatgpt} }
    \label{fig:listwiseprompt}
\end{figure}
The advantage of a listwise prompt over a pairwise prompt is that it is more efficient: a listwise prompt generates a ranking for all documents at once, compared to a pairwise prompt that needs to be repeated for each pair of documents. 
\item \textbf{Setwise}~\citep{zhuang2024setwise}: asking the LLM to select the most relevant document from a set. Give a query and a set of retrieved documents, the model is asked to output the DocID of the most relevant document. The logits or probabilities assigned by the LLM to each DocID are used to rank the documents. Compared to listwise, the setwise approach is faster because it only generates a single output instead of a full list. 
\end{itemize}

LLM output is known to be sensitive to changes in prompts, and this is no different for ranking prompts. \citet{sun2025investigation} show that the specific instructions in the prompt have a large effect on the success of LLM-based rankers. They pointed out that with an optimized prompt, pointwise ranking can be as good as listwise ranking.

Irrespective of the prompt used, LLM-based re-ranking has the advantage over cross-encoder re-ranking that it does not require supervised learning on a training set of query-document pairs with relevance assessments, but is completely unsupervised. This makes it very useful in domain-specific settings where no relevance assessments are available.

\paragraph{Model-based Generative IR.}

The idea of training a generative model instead of a discriminative model to predict relevant DocIDs was first introduced by \citet{tay2022transformer} in the form of a model called Differentiable Search Index (DSI). DSI learns a sequence-to-sequence model (T5) that takes queries as input and generates relevant document IDs (docIDs). In follow-up papers, proposals were made to replace numeric docIDs by meaningful identifiers~\citep{lee2023glen,sun2023learning}. \begin{figure}[t]
    \centering    \includegraphics[width=0.6\linewidth]{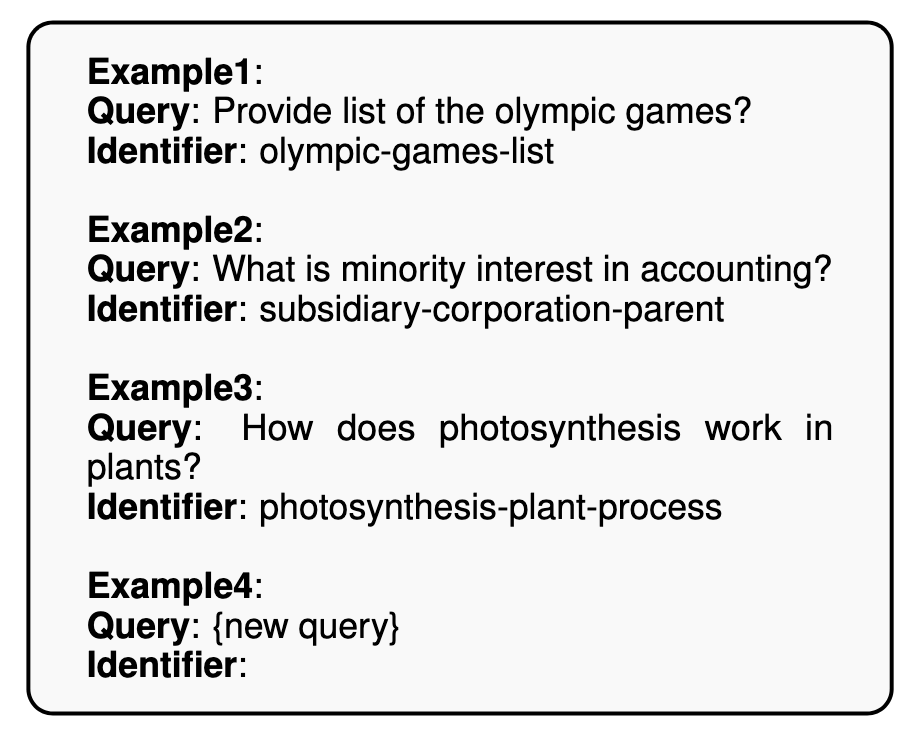}
    \caption{Prompt used for indexing and retrieval in few-shot generative IR~\citep{askari2024generative}.}
    \label{fig:fewshotgenIR}
\end{figure} 
Model-based generative IR is motivated by the simplification of the pipeline by integrating the indexing, retrieval and ranking into a single architecture. One disadvantage though is that training the sequence-to-sequence model for the content-to-docID task is computationally heavy and data-hungry, which to a large extent disqualifies the approach for domain-specific contexts that often lack sufficient training data and take place in low-resource settings. One potential solution to solving these issues is replacing a fine-tuned sequence-to-sequence model by a generative LLM with few-shot learning (see Figure~\ref{fig:fewshotgenIR})~\citep{askari2024generative}. More work in this direction is still needed to find out if it will be adopted as one of the future paradigms for IR.

\chapter{Conclusions}
\label{chap:conclusions}

We reflect on the content of this survey, first by providing a summary and main findings, then discussing implications and limitations. We conclude with an outline of our research agenda, focussing on the challenges listed in Chapter~\ref{chap:challenges}.

\section{Summary and main findings}
We have discussed domain-specific Information Retrieval from four different aspects: 
\begin{itemize} \item \textbf{The domain: tasks, users and content} (Chapter~\ref{chap:domain-specific_IR}). We focused on four domains in which most of the academic research is situated: Legal, Health and Biomedical, Academic, and Patents. We have discussed tasks, applications, and datasets for each of them. In most of these domains, it often is a professional (expert) performing the search, but it is certainly not the case that domain-specific search equals professional search; With the increasing democratization and accessibility of information to laypeople, it is common for general web search users to search for medical or legal information online. The legal domain is a clear example where professional tasks, such as case law retrieval or eDiscovery, are very different from layperson tasks such as legal community Question Answering. It is important that this diversity is reflected in test collections and in IR research.
\item \textbf{The challenges} (Chapter~\ref{chap:challenges}). We discussed three challenges that are commonly mentioned as challenges in domain-specific IR: (1) long documents and long queries, (2) complex and specific user needs, and (3) session-based and exploratory search. In addition, we addressed three challenges that are common for IR in general, that we consider of high importance for domain-specific IR, and that have their own particular aspects when it comes to specific domains: (4) reproducibility, (5) bias and fairness, and (6) evaluation. We come back to these six challenges in Section~\ref{sec:agenda}, where we outline our research agenda.
\item \textbf{The solutions: models and methods} (Chapter~\ref{chap:retrieval_models_methods}). Forming the core of our survey, Chapter~\ref{chap:retrieval_models_methods} provides an overview of retrieval models and methods that are relevant to domain-specific problems. Starting from the general set-up of two-stage retrieval, we followed up with methods for domain transfer and query processing methods. We then discussed methods that directly relate to the challenges presented in Chapter~\ref{chap:challenges}: dealing with long queries and long documents, developing more efficient methods, method development for domain-specific conversational search, and approaches for data augmentation. Although each of these developments are directly motivated by domain-specific challenges, they are also relevant for general domain web search and for the IR field at large. 
\item \textbf{The current landscape: domain-specific IR in the age of LLMs} (Chapter~\ref{chap:LLMs}). The relation between IR and LLMs has gained substantial attention since 2022. While the general public has increasingly embraced LLM-based chatbots for information finding, IR remains important for grounding generated responses in sources. At the same time, the generative AI paradigm has also has shown its potential for the development of new, flexible, potentially more efficient retrieval models, especially in the context of limited labelled data. 
\end{itemize}

\section{Implications} 
The rise of generative AI \revision{ has led to the emergence of} new paradigms for search and indexing data: RAG-based conversational search, prompt-based generative IR and model-based generative IR (Section~\ref{sec:genLLMs}). It has even been argued that long-context LLMs may make traditional indexes unnecessary, because the whole corpus can be fed into the prompt~\citep{lee2024can}. 

We argue that changing the paradigm for indexing and retrieval does not invalidate IR, because many of the tasks and problems that are studied in IR continue to be relevant, irrespective of the methods or tools for indexing and retrieval: 
\begin{itemize}[noitemsep,nolistsep] 
    \item the right knowledge corpus needs to be identified for a query; 
    \item a ranking is needed to provide the most relevant information to the user (or LLM); 
    \item a distinction is needed between navigational, informational, exploratory, and high-recall tasks; 
    \item a solution is needed for information being spread in long and over multiple documents; 
    \item diversification is necessary when handling ambiguous queries; 
    \item transparency is needed for the user to show why the retrieved documents are considered relevant (e.g. snippets with highlighting); 
    \item evaluation of the results is needed, depending on the task; and, finally, 
    \item  we need the user to formulate the queries and select the relevant information. 
\end{itemize}

This is all even more important when it comes to \textit{domain-specific} problems. Research into domain-specific IR can set examples and provide case studies for method development in the IR field at large: problems that seem specific to certain domains at first sight (such as complex user queries or session-based search) occur in many other IR contexts as well, especially because regular web search users also commonly engage in complex search tasks.

\section{Limitations} \label{sec:limitations} 

Information Retrieval and its related fields (Natural Language Processing, Machine Learning) are developing fast. By definition, a survey of models and methods that is written in 2025 will be partially outdated in 2027. We have therefore tried to focus on the long-standing challenges and model-agnostic research topics. This includes the domain aspects in Chapter~\ref{chap:domain-specific_IR} and the challenges discussed in Chapter~\ref{chap:challenges}. The majority of methods discussed in Chapter~\ref{chap:retrieval_models_methods} are model-agnostic and therefore will still be relevant in the coming years. The relations between IR and LLMs (Chapter~\ref{chap:LLMs}) are subject to fast development, and 2023 papers may no longer be relevant in 2027. The overall argument of the importance of sources and source attributions, however, will remain relevant and may even become more important as search behaviour is moving more and more from traditional ranking systems to conversational, LLM-driven systems. Especially in domain-specific contexts (legal, medical, academic), it is important that the output of search systems is verifiable and factually correct.

In terms of scope, we have originally excluded Question Answering (Section~\ref{sec:scope}), with the addition that ``the retrieval step could be used as a first step in Question Answering.'' In the course of the survey writing, this topic has come back a couple of times, because of the increased interest in conversational search and retrieval-augmented generation in domain-specific contexts. In our discussion of the literature on QA and RAG (e.g. section~\ref{sec:RAG}), we have focused on the retrieval aspects of these topics. 

A limitation of this survey, and of the field as a whole, is the focus on English; the large majority of papers we discussed worked on models and methods for English domain-specific retrieval. In the scientific and biomedical contexts, this is reasonable since the majority of scientific literature is written in English. In the arts and humanities, briefly discussed in Section~\ref{sec:academic}, we have mentioned a number of papers working in non-English European contexts, but we have not explicitly provided an overview of work along the dimension of language. In the legal and patent domains, other languages than English are definitely relevant, because of the different law systems and national patent offices. We have discussed this in the respective sections (\ref{sec:legal} and \ref{sec:patents}).

\section{Our research agenda} \label{sec:agenda} 
The challenges discussed in Chapter~\ref{chap:challenges} form the basis for our research agenda for domain-specific IR:

\paragraph{(1) Dealing with long documents} The challenge of long documents remains relevant in the age of LLMs: just like humans, LLMs can suffer from information overload and struggle to identify the most relevant information in long contexts~\citep{liu2024lost}. In the implementation of RAG-systems, therefore, mostly only the top-5 passages are sent to the generator, and there is only a small positive correlation between relevance of the retrieved passages and the utility of the output~\citep{tian2025relevance}. In domain-specific contexts, documents often are much longer than passage-length nuggets and information is spread throughout the whole document (see Section~\ref{sec:legal}). For that reason, domain-specific RAG poses challenges that need attention from the scientific community. 

\paragraph{(2) Dealing with complex and specific user needs} With the increased use of conversational search interfaces with instruction-tuned LLMs, users have higher expectations of search engines to answer complex questions. LLMs can be prompted to answer complex questions that require multiple reasoning steps, a task that is referred to as Multi-Hop QA~\citep{mavi2024multi,yang2018hotpotqa}. In each step of the multi-hop process, a partial answer is found, and the partial answers need to be combined into a final answer to the question. Although there is a small number of medical and scientific benchmark sets for multi-hop QA available~\citep[Table 3.1]{mavi2024multi}, there is still much to gain in terms of methods development for these and other domains~\citep[Table 6.1]{mavi2024multi}. In that light it is interesting that \citet[Section 9.3]{mavi2024multi} warn that new datasets for multi-hop QA should not depend on knowledge bases (KB) and instead be in the open-domain setting. This raises the question: what makes complex retrieval and question answering in domains like legal, health, or science, different from complex retrieval and question answering in the open-domain? And is there even a real difference, or can modern techniques for RAG-based complex question answering be applied in either domain? Only once these questions are answered, the community can focus on developing methods that solve the problems specific to the domains. 

\paragraph{(3) Models for session-based and exploratory search} The IR field has long recognized the significance of session-based evaluation~\citep{belkin2012whole,wicaksono_modeling_2021}: Users can be guided much better in exploratory search, e.g. by providing query suggestions, summarization, and transparency about query interpretation. Experimental assessments however still mostly rely on query-level test collections~\citep{lin2022pretrained}. For future research, advancing session-based retrieval models will be crucial for optimizing retrieval effectiveness, especially in the context of conversational interfaces. While the TREC Session Track~\citep{carterette2014overview} provided valuable insights, these principles should be extended to domain-specific (and even personalized) settings. Also in evaluation of retrieval effectiveness, the session context should play a bigger role: which part of the task was solved in previous query turns, and which subset of the provided information in the current turn is novel and useful?

\paragraph{(4) Reproducibility} deserves attention because in domain-specific problems, not everyone has access to the same data. The data may be proprietary, enterprise-level, personal, or behind publisher paywalls with different access levels. With compute-heavy transformer models, differences in compute availability also play a role: not everyone can train their models on full datasets and with equal number of epochs; the sheer size of these models sometimes forces researchers to reproduce the results with smaller models (e.g. an 8B Llama model instead of a 70B Llama model). In addition, generative models are more difficult to reproduce than discriminative models because their output may vary from run to run. In this light, it is \revision{commendable} that IR conferences have reproducibility paper tracks. In a somewhat different angle of reproducibility, motivated by the importance of controlled search strategies, reproducibility of search results by diverse users (e.g. two colleagues from the same academic field) or user groups (e.g. medical professionals vs. laypeople) is a relevant topic that has a direct connection to domain-specificity.

\paragraph{(5) Bias and fairness} Societal bias has received limited attention in IR until now. We have described some work that proposes metrics for measuring bias in a particular dimension (e.g. gender); the work that proposes methods to mitigate bias in rankings is even more limited. The role of bias in RAG systems has received attention recently, but it is still an open question what the role of bias is in domain-specific IR and RAG. This deserves attention in the coming years.

\paragraph{(6) Evaluation} Top-$k$ rankings are commonly evaluated for relevance with retrieval metrics such as Precision@k, Recall@k, and Mean Average Precision (MAP). Relevance is context-dependent, however, and not synonymous with utility: a document in itself may be relevant but have low utility in the context of a full search session for a complex user need; a document that is relevant for one user may not be relevant to another user; and a document that is relevant to a human may not be useful by an LLM in a RAG context. Therefore, evaluation beyond relevance and beyond top-$k$ rankings is important. 

\chapter*{Acknowledgements}
This work was partly supported by the EU Horizon 2020 ITN/ETN on Domain Specific Systems for Information Extraction and Retrieval -- DoSSIER (H2020-EU.1.3.1., ID: 860721).


\printbibliography

\end{document}